\documentclass{article}

\scrollmode
\usepackage{amsmath}
\usepackage{amsfonts}
\usepackage{amssymb}
\usepackage{latexsym}
\usepackage{stmaryrd}
\usepackage{array}
\usepackage{exscale}
\newcommand{\nc}{\newcommand}
\newcommand{\ol}{\overline}

\newcommand{\es}{\emptyset}
\newcommand{\sm}{\setminus}
\newcommand{\ve}{\varepsilon}
\newcommand{\vp}{\varphi}

\newcommand{\bc}{\bigcup}

\newcommand{\Lra}{\Leftrightarrow}

\newcommand{\Ra}{\Rightarrow}

\newcommand{\ra}{\rightarrow}

\newcommand{\la}{\leftarrow}
\newcommand{\lra}{\leftrightarrow}

\newcommand{\sse}{\subseteq}

\newcommand{\spe}{\supseteq}
\newcommand{\fa}{\forall}
\newcommand{\ex}{\exists}
\newcommand{\mr}{\mathrm}
\newcommand{\mc}{\mathcal}
\newcommand{\mf}{\mathfrak}

\newcommand{\DMO}{\DeclareMathOperator}
\newcommand{\DST}{\displaystyle}

\newcommand{\ZZ}{\mathbb{Z}}
\newcommand{\NN}{\mathbb{N}}
\newcommand{\NNZ}{\NN_0}

\newcommand{\RR}{\mathbb{R}}

\newcommand{\PP}{\mathbb{P}}

\newcommand{\aru}{\ar @{-}} 
\usepackage{listings}
\newcommand{\inl}[1]{\lstinline$#1$}
\newcommand{\und}{{\:\wedge\:}} 
\newcommand{\oder}{{\:\vee\:}} 
\newcommand{\mb}{{\:|\:}} 
\newcommand{\set}[1]{\{ #1 \}}
\newcommand{\setb}[1]{\big \{ \, #1 \, \big \}}
\nc{\simlvi}[1]{\!\sim_{#1}}
\DeclareMathOperator{\rstr}{|} 
\nc{\apprel}[3]{{#1}(#2)_{(#3)}} 
\nc{\cmpli}[1]{\complement^1_{#1}} 
\nc{\cmplzi}[1]{\complement^0_{#1}} 
\nc{\cmplzoi}[1]{\complement^*_{#1}} 

\nc{\zf}{\mr{ZF}}
\nc{\zfmf}{\zf^0} 
\nc{\zfc}{\mr{ZFC}}
\nc{\zfcmf}{\zfc^0} 
\nc{\bst}{\mr{BST}} 
\newcommand{\tb}[2]{\set{#1, \dots, #2}} 
\providecommand{\abs}[1]{\lvert #1 \rvert} 
\providecommand{\norm}[1]{\lVert #1 \rVert} 
\newcommand{\trans}[1]{#1^{\hspace{0.05em}\mr{t}}} 
\makeatletter
\DeclareRobustCommand{\genericinterval}[2]{%
  \@ifstar{\genericinterval@star{#1}{#2}}{\genericinterval@nostar{#1}{#2}}}
\newcommand{\genericinterval@star}[4]{\mathopen{}\mathclose{\left#1#3,#4\right#2}}
\newcommand{\genericinterval@nostar}[4]{\mathopen{#1}#3,#4\mathclose{#2}}

\makeatother
\nc{\untit}[2]{{#1}^{#2 \downarrow}} 
\nc{\obit}[2]{{#1}^{#2 \uparrow}} 
\nc{\inzEKi}[1]{\mc{I}^{\mr{V}}_{#1}}

\nc{\inzKEi}[1]{\mc{I}^{\mr{E}}_{#1}}

\nc{\adjEi}[1]{\mc{A}^{\mr{V}}_{#1}}

\nc{\BD}[1]{{#1}\text{-}\mr{BD}}

\DeclareMathOperator{\pote}{\PP_f} 
\nc{\konv}[2]{{#1}[{#2}]} 
\nc{\actpres}[1]{\Phi_{#1}} 
\nc{\Prim}{\mc{PR}} 

\nc{\sselr}{\sse^{\mapsto}}
\nc{\sserl}{\sse^{\mapsfrom}}
\nc{\spelr}{\spe^{\mapsto}}
\nc{\sperl}{\spe^{\mapsfrom}}
\nc{\ball}[1]{\mr{B}^{#1}} 
\nc{\oball}[1]{\breve{\mr{B}}^{#1}} 
\nc{\pball}[1]{\dot{\mr{B}}^{#1}} 
\nc{\prr}[1]{\dot{\RR}^{#1}} 
\nc{\sph}[1]{\mr{S}^{#1}} 
\nc{\ssim}[1]{s\sigma_{#1}} 
\nc{\koerper}[1]{\norm{#1}}
\nc{\Ccovdim}{\mc{CD}}
\nc{\Cinddim}{\mc{SID}}

\nc{\CInddim}{\mc{LID}}

\DeclareMathOperator{\diffop}{D} 
\DeclareMathOperator*{\diffoplimit}{D} 
\nc{\diffopc}[1]{\sideset{_{#1}}{}\diffoplimit} 
\nc{\diffopp}[1]{\diffop_{#1}} 
\nc{\diffopcp}[2]{\sideset{_{#2}}{_{#1}}\diffoplimit} 
\nc{\meanH}[2]{\mf{M}_{#1,#2}} 
\nc{\emean}[2]{\mf{M}_{\exp_{#1},#2}} 
\DeclareMathOperator{\mor}{Mor}
\DeclareMathOperator{\Hom}{Hom} 
\nc{\autoerw}[1]{\mr{Aut}^{#1}} 
\nc{\komma}[2]{(#1 \downarrow #2)} 
\nc{\Kmat}{\mf{MAT}} 
\nc{\Khmat}{\mf{HMAT}} 
\nc{\homfun}[1]{\mor_{#1}(-_1,-_2)} 
\nc{\homfunae}[1]{\mor_{#1}(-_1)} 
\nc{\homfunbe}[1]{\mor_{#1}(-_2)} 
\nc{\homfunxy}[3]{\mor_{#1}(#2(-_1), #3(-_2))}
\nc{\homfunx}[2]{\mor_{#1}(#2(-_1), -_2)}
\nc{\homfuny}[2]{\mor_{#1}(-_1, #2(-_2))}
\nc{\homfuna}[2]{\mor_{#1}(#2, -)} 
\nc{\homfunb}[2]{\mor_{#1}(-, #2)} 
\nc{\hhomfuna}[2]{\Hom_{#1}(#2, -)} 
\nc{\hhomfunb}[2]{\Hom_{#1}(-, #2)} 
\newcommand{\Va}{\mc{V\hspace{-0.1em}A}}

\newcommand{\Lit}{\mc{LIT}}
\newcommand{\Cl}{\mc{CL}}
\newcommand{\Cls}{\mc{CLS}}

\newcommand{\Pcls}[1]{#1\mbox{--}\Cls}

\newcommand{\Pass}{\mc{P\hspace{-0.32em}ASS}}
\newcommand{\epa}{\pab{}} 

\newcommand{\Sat}{\mc{SAT}}

\newcommand{\Usat}{\mc{USAT}}

\newcommand{\Musat}{\mc{M\hspace{0.8pt}U}} 
\newcommand{\Musati}[1]{\Musat_{\!#1}} 
\newcommand{\Smusat}{\mc{S}\Musat} 
\newcommand{\Smusati}[1]{\Smusat_{\!#1}}

\nc{\Clsoo}{\Cls^{1,1}} 
\DeclareMathOperator{\lit}{lit}
\DeclareMathOperator{\var}{var}

\newcommand{\Clash}{\mc{HIT}} 

\newcommand{\Uclash}{\mc{U}\Clash} 
\newcommand{\Uclashi}[1]{\Uclash_{\!\!#1}}

\newcommand{\Ho}{\mc{HO}} 

\DeclareMathOperator{\res}{\diamond} 

\DeclareMathOperator{\comp}{Comp} 
\DeclareMathOperator{\compex}{\comp_{ER}} 
\DeclareMathOperator{\comptex}{\comp_{ER}^*} 
\DeclareMathOperator{\compr}{\comp_R} 
\DeclareMathOperator{\comptr}{\comp_{R}^*} 
\DMO{\premr}{F} 
\DMO{\concr}{C} 
\DMO{\allcr}{\widehat{F}} 
\DMO{\semspace}{ss} 
\DMO{\resspace}{rs} 
\DMO{\treespace}{ts} 

\DeclareMathOperator{\hardness}{hd}
\DMO{\thardness}{thd} 
\DMO{\phardness}{phd} 
\DMO{\whardness}{whd} 
\DMO{\dep}{dep} 
\DMO{\hts}{hs} 
\newcommand{\php}{\mathrm{PHP}}
\newcommand{\fphp}{\mathrm{FPHP}} 
\newcommand{\ophp}{\mathrm{OPHP}} 
\newcommand{\ofphp}{\mathrm{BPHP}} 
\newcommand{\ephp}{\mathrm{EPHP}} 
\DeclareMathOperator{\rt}{rt} 
\DeclareMathOperator{\nds}{nds} 
\DeclareMathOperator{\lvs}{lvs} 
\DeclareMathOperator{\nlvs}{\#lvs} 
\DeclareMathOperator{\height}{ht}

\DeclareMathOperator{\peb}{peb} 
\newcommand{\pab}[1]{\langle #1 \rangle}
\newcommand{\pao}[2]{\langle #1 \ra #2 \rangle}

\nc{\bth}[1]{\langle{#1}\rangle} 
\DMO{\rsub}{r_S} 
\DMO{\rk}{r} 
\DMO{\rki}{r_{\infty}} 
\nc{\rslur}{\xrightarrow{\text{SLUR}}} 
\nc{\rslurs}{\rslur_{\!*}} 
\DMO{\slur}{slur} 
\nc{\Slur}{\mc{SLUR}} 
\nc{\rkslur}[1]{\xrightarrow{\text{SLUR}_{#1}}} 
\nc{\rkslurs}[1]{\rkslur{#1}_{\!*}} 
\nc{\Altsluri}[1]{\Slur(#1)}
\nc{\Altslurstari}[1]{\Slur\text{\textasteriskcentered}(#1)}
\nc{\Canoni}[1]{\mr{CANON}(#1)}
\nc{\rkslurstar}[1]{\xrightarrow{\text{SLUR\textasteriskcentered}#1}} 
\nc{\rkslursstar}[1]{\rkslurstar{#1}_{\!*}} 
\DMO{\slurstar}{\slur\!\text{\textasteriskcentered}}
\nc{\Urefc}{\mc{UC}}
\nc{\Propc}{\mc{PC}}
\nc{\Wrefc}{\mc{WC}} 
\DeclareMathOperator{\wid}{wid} 

\DeclareMathOperator{\vdeg}{vd} 
\DeclareMathOperator{\minvdeg}{\mu\!\vdeg} 
\DMO{\varmvd}{\var_{\minvdeg}} 
\DMO{\nfc}{fc} 
\DMO{\maxnfc}{\nu\!\nfc} 
\nc{\svbf}{\mc{VB}} 
\nc{\svbfs}{\mc{VB}^*} 
\DMO{\potp}{pp} 
\DMO{\potprec}{NM} 
\DMO{\minnonmer}{\mu{}nM} 
\DMO{\varsing}{\var_s} 
\DMO{\varosing}{\var_{1s}} 
\DMO{\varnosing}{\var_{\neg1s}} 
\nc{\Musatns}{\Musat'} 
\nc{\Musatnsi}[1]{\Musati{#1}'}
\nc{\Smusatns}{\Smusat'} 
\nc{\Smusatnsi}[1]{\Smusati{#1}'}
\nc{\Uclashns}{\Uclash'} 
\nc{\Uclashnsi}[1]{\Uclashi{#1}'}
\nc{\tsdp}{\xrightarrow{\text{sDP}}}
\nc{\tsdps}{\tsdp_{\!*}}
\nc{\tosdp}{\xrightarrow{\text{1sDP}}}
\nc{\tosdps}{\tosdp_{\!*}}
\DMO{\sdp}{sDP} 
\DMO{\osdp}{sDP_1} 
\nc{\cflmusat}{\mc{CF}\Musat} 
\nc{\cflmusati}[1]{\mc{CF}\Musati{#1}}
\nc{\cflimusat}{\mc{CFI}\Musat} 
\DMO{\sNF}{sNF} 
\DMO{\eqp}{eqp} 
\DMO{\sgp}{sp} 
\DMO{\singind}{si} 
\DMO{\osingind}{si_1} 
\DMO{\shyp}{svh} 
\DMO{\sdph}{ssh} 
\DMO{\msdph}{mss} 
\DMO{\osdph}{ssh_1} 
\DMO{\mosdph}{mss_1} 
\DMO{\mps}{mps} 
\DMO{\purec}{puc} 
\DMO{\doping}{D}
\DeclareMathOperator{\primec}{prc} 
\nc{\glue}[4]{\mr{glue}((#1,#2), (#3,#4))} 
\DMO{\fvdglue}{\boxplus} 
\nc{\gluea}[3]{#1 \boxplus_{#3} #2} 
\DMO{\frl}{fl} 
\nc{\Con}{\mr{Con}}
\nc{\Log}{\mr{Log}}
\nc{\Lin}{\mr{Lin}}
\nc{\Pol}{\mr{Pol}}
\nc{\ExL}{\mr{ExL}}
\nc{\ExP}{\mr{ExP}}
\nc{\CTime}{\mr{CTime}}
\nc{\CSpace}{\mr{CSpace}}
\nc{\LTime}{\mr{LTime}}
\nc{\LSpace}{\mr{L}}
\nc{\NLSpace}{\mr{NL}}
\nc{\LinTime}{\mr{LinTime}}
\nc{\LinSpace}{\mr{LinSpace}}
\nc{\PTime}{\mr{P}}
\nc{\PSpace}{\mr{PSpace}}
\nc{\Np}{\mr{NP}}
\nc{\Conp}{\text{coNP}}
\nc{\NPSpace}{\mr{NPSpace}}
\nc{\CoNPSpace}{\mr{coNPSpace}}
\nc{\ELTime}{\mr{ELTime}}
\nc{\ELSpace}{\mr{ELSpace}}
\nc{\EPTime}{\mr{EPTime}}
\nc{\EPSpace}{\mr{EPSpace}}
\nc{\NEPTime}{\mr{NEPTime}}
\nc{\polydelta}[1]{\Delta_{#1}^{\mr P}}
\nc{\polypi}[1]{\Pi_{#1}^{\mr P}}
\nc{\polysigma}[1]{\Sigma_{#1}^{\mr P}}
\nc{\Ph}{\mr{PH}}

\nc{\Dp}{D^P}
\nc{\PllC}[2]{{\text{$\mr{PT}$/$\mr{WK}$}(#1, #2)}} 
\nc{\Nc}{\mr{NC}}
\nc{\Nci}[1]{\Nc^{#1}}
\nc{\Ac}{\mr{AC}}
\nc{\Aci}[1]{\Ac^{#1}}
\nc{\pmodpoly}{P / \mathrm{poly}}
\nc{\Wh}[1]{\mr{W}[#1]} 
\nc{\Rl}{\mr{RL}}
\nc{\coRl}{\mr{coRL}}
\nc{\Rp}{\mr{RP}}
\nc{\coRp}{\mr{coRP}}
\nc{\Zpp}{\mr{ZPP}}
\nc{\Bpp}{\mr{BPP}}
\nc{\Pp}{\mr{PP}}
\nc{\Reach}{\mr{STCON}} 
\nc{\Undreach}{\mr{USTCON}} 
\nc{\Pcol}[2]{\mr{COL}(#1,#2)} 
\nc{\Pscol}[2]{\mr{SCOL}(#1,#2)} 
\nc{\Psorcol}[2]{\mr{SORCOL}(#1,#2)} 
\DMO{\slp}{slp}
\nc{\Mss}{\mr{MSS}}
\nc{\Key}{\mr{KEY}}
\nc{\Keyi}[1]{\Key_{\!#1}}
\nc{\Nbmss}{N_{\mr{bm}}} 
\nc{\Nbkey}{N_{\mr{bk}}} 
\nc{\Rnb}{N_{\mr{b}}}
\nc{\Rnk}{N_{\mr{k}}}
\nc{\Rnr}{N_{\mr{r}}}

\nc{\Byte}{\mr{B}[8]}
\nc{\QByte}{\mr{B}[4,8]}
\nc{\KByte}{\mc{B}} 
\nc{\RQByte}{\mc{QB}} 

\nc{\ramz}[3]{\mr{ram}_{#1}^{#2}(#3)} 
\nc{\waez}[2]{\mr{vdw}_{#1}(#2)} 
\nc{\gtz}[2]{\mr{grt}_{#1}(#2)} 
\nc{\pdwaez}[2]{\mr{vdw}_{#1}^{\mr{pd}}(#2)} 
\nc{\absfeh}[1]{\delta_{#1}} 
\nc{\relfeh}[1]{\ve_{#1}} 
\usepackage{theorem} 
\usepackage[driverfallback=hypertex]{hyperref}
\newtheorem{defi}{Definition}[section]
\newtheorem{lem}[defi]{Lemma}
\newtheorem{thm}[defi]{Theorem}
\newtheorem{corol}[defi]{Corollary}

\newtheorem{conj}[defi]{Conjecture}

\newtheorem{examp}[defi]{Example}

\newtheorem{quest}[defi]{Question}

\theorembodyfont{\rmfamily}

\theorembodyfont{}
\newenvironment{prf}{\noindent\textbf{Proof:}\;}{\par\noindent\ignorespacesafterend}

\newcommand{\Qed}{\hfill $\square$}
\newcounter{dDef} 

\newcounter{dLem} 

\newcounter{dThm} 

\newcounter{dPro} 

\newcounter{Beispielzaehler}

\nc{\bm}{\boldmath}
\nc{\bmm}[1]{\mbox{\bm$\DST #1$}}
\nc{\mi}[1]{\bmm{\mathrm{(#1):}} \quad}
\usepackage[active]{srcltx}

\usepackage{a4}
\usepackage[all]{xy}
\usepackage{enumerate}
\usepackage{xr}

\usepackage{xspace}
\newcommand{\Pudlak}{Pudl\'{a}k\xspace}

\begin{document}

\title{Hardness measures and resolution lower bounds}

\author{Olaf Beyersdorff\\
  \href{http://www.engineering.leeds.ac.uk/computing/}{School of Computing}\\
  \href{http://www.leeds.ac.uk/}{University of Leeds}\\
  {\small \url{http://www.engineering.leeds.ac.uk/people/computing/staff/o.beyersdorff}}
  \and
  Oliver Kullmann\\
  \href{http://www.swan.ac.uk/compsci/}{Computer Science Department}, \href{http://www.swan.ac.uk/science/}{College of Science}\\
  \href{http://www.swan.ac.uk/}{Swansea University, UK}\\
  {\small \url{http://cs.swan.ac.uk/~csoliver}}
}

\maketitle

\begin{abstract}
  Various ``hardness'' measures have been studied for resolution, providing theoretical insight into the proof complexity of resolution and its fragments, as well as explanations for the hardness of instances in SAT solving. In this report we aim at a unified view of a number of hardness measures, including different measures of width, space and size of resolution proofs. We also extend these measures to all clause-sets (possibly satisfiable).

  One main contribution is a unified game-theoretic characterisation of these measures. We obtain new relations between the different hardness measures. In particular, we prove a generalised version of Atserias and Dalmau's result on the relation between resolution width and space from \cite{AD2002J}.

  As an application, we study hardness of PHP and variations, considering also satisfiable PHP. Especially we consider EPHP, the extension of PHP by Cook (\cite{Co76}) which yields polynomial-size resolution refutations. Another application is to XOR-principles.
\end{abstract}

\tableofcontents

\section{Introduction}
\label{sec:intro}

Arguably, resolution is the best understood among all propositional proof system, and at the same time it is the most important one in terms of applications. To understand the complexity of resolution proofs, a number of \emph{hardness measures} have been defined and investigated. Historically the first and most studied measure is the \emph{size of resolution proofs}, with the first lower bounds dating back to Tseitin \cite{Ts68} and Haken \cite{Hak85}.
A number of ingenious techniques have been developed to show lower bounds for the size of resolution proofs, among them feasible interpolation \cite{Kra97} which applies to many further systems. In their seminal paper \cite{SW98J}, Ben-Sasson and Wigderson showed that resolution size lower bounds can be very elegantly obtained by showing lower bounds to the \emph{width} of resolution proofs. Indeed, the discovery of this relation between width and size of resolution proofs was a milestone in our understanding of resolution. Around the same time \emph{resolution space} was investigated, and first lower bounds were obtained \cite{To99,ET99b}. The primary method to obtain lower bounds on resolution space is based on width, and the general bound was shown in the fundamental paper by Atserias and Dalmau \cite{AD2002J}. Since then the relations between size, width and space have been intensely investigated, resulting in particular in sharp trade-off results \cite{BBI12,BN11,Nordstrom2012PebbleProofTimeSpaceSurvey,NordstroemHastad2013SpaceLength}. Space of tree resolution has also been investigated in \cite{Ku99b,Ku000,Ku00g}, called ``hardness'' and with an algorithmic focus (closely related to size of tree resolution, as shown in \cite{Ku99b}; one can also say ``tree-hardness''), together with a generalised form of width, which we call ``asymmetric width'' in this report.

One of the prime motivations to understand these measures is their close correspondence to SAT solving (see \cite{2008HandbuchSAT} for general information). In particular, resolution size and space relate to the running time and memory consumption, respectively, of executions of SAT solvers on unsatisfiable instances. However, size and space are not the only measures which are interesting with respect to SAT solving, and the question what constitutes a good hardness measure for practical SAT solving is a very important one (cf.\ \cite{AnsoteguiBonetLevyManya2008Hardness,JaervisaloNordstroem2012PracticalHardness} for discussions).

The aim of this report is to review different hardness measures defined in the literature, and to provide \emph{unified characterisations} for these measures in terms of Prover-Delayer games and sets of partial assignments satisfying some consistency conditions. These unified characterisations allow elegant proofs of basic relations between the different hardness measures. Unlike in the works \cite{BBI12,BN11,Nordstrom2012PebbleProofTimeSpaceSurvey}, our emphasis is here not on trade-off results, but on \emph{exact relations} between the different measures.

For a clause-set $F$ (possibly satisfiable) we consider the following measures, related to resolution proofs of prime implicates (clauses which are logically entailed):
\begin{description}
\item[Size] (or ``shape'')
  \begin{itemize}
  \item the \textbf{depth} $\dep(F)$, the maximal depth (introduced in \cite{Urquhart2011Depth} for unsatisfiable clause-sets)
  \item the \textbf{hardness} $\hardness(F)$, the maximal level of nested input resolution (introduced for general clause-sets in \cite{GwynneKullmann2012SlurSOFSEM,GwynneKullmann2012SlurJ}, based on \cite{Ku99b,Ku00g})
  \end{itemize}
   needed to derive any prime implicate.
\item[Width] (of clauses)
  \begin{itemize}
  \item the \textbf{symmetric width} $\wid(F)$, the maximal clause-length (introduced in \cite{SW98J}, based on \cite{CEI96})
  \item the \textbf{asymmetric width} $\whardness(F)$, the maximal minimal parent-clause length (introduced for general clause-sets in \cite{GwynneKullmann2012SlurJ,GwynneKullmann2013GoodRepresentationsI,GwynneKullmann2013GoodRepresentationsII}, based on \cite{Kl93,Ku99b,Ku000,Ku00g})
  \end{itemize}
   needed to derive any prime implicate.
\item[Space] (number of clauses needed to be considered at once)
  \begin{itemize}
  \item \textbf{semantic space} $\semspace(F)$ (introduced in \cite{AlekhnovichBRW02} for unsatisfiable clause-sets)
  \item \textbf{resolution space} $\resspace(F)$ (introduced in \cite{KlLe94,KlLe99,To99,ET99b} for unsatisfiable clause-sets)
  \item \textbf{tree-re\-so\-lu\-tion space} $\treespace(F)$ (introduced in \cite{To99} for unsatisfiable clause-sets)
  \end{itemize}
  needed to derive any prime implicate.
\end{description}

\subsection{Game-theoretic methods}

Using games has a long tradition in proof complexity, as they provide intuitive and simplified methods for lower bounds in resolution, e.g.\ for Haken's exponential bound for the pigeonhole principle in dag-like resolution \cite{Pud00}, or the optimal bound in tree resolution \cite{BeyersdorffGalesiLauria2010Treelike}, and even work for very strong systems \cite{BH10}. Inspired by the Prover-Delayer game of \Pudlak and Impagliazzo \cite{PI2000}, we devise a game that characterises the hardness measure $\hardness(F)$, but in contrast to \cite{PI2000} also works for satisfiable formulas (Theorem~\ref{thm:hd=PI}). We then explain a more general game allowing the Prover to also \emph{forget} some information. This game tightly characterises the asymmetric width hardness $\whardness(F)$ (Theorem~\ref{thm:whdspiel}); and restricting this game by disallowing forgetting yields the $\hardness$-game (Lemma~\ref{lem:althdg}).

\subsection{Consistency notions}

Characterisations by partial assignments provide an alternative combinatorial description of the hardness measures. In \cite{AD2002J} such a characterisation is obtained for the symmetric width $\wid(F)$. Taking this as a starting point, we devise a hierarchy of consistency conditions for sets of partial assignments which serve to characterise  asymmetric width $\whardness(F)$ ($k$-consistency, Theorem~\ref{thm:methodwhdlb}), hardness $\hardness(F)$ (weak $k$-consistency), and depth $\dep(F)$ (very weak $k$-consistency).

\subsection{Relations between these measures}

We obtain a generalised version of Atserias and Dalmau's connection between width and resolution space from \cite{AD2002J}, where we replace symmetric width by the stronger notion of asymmetric width (handling long clauses now), and resolution space by the tighter semantic space (Theorem~\ref{thm:resspace}). The full picture is presented in the following diagram, where $F \in \Cls$ has $n$ variables, minimal clause length $p$, and maximal length $q$ of necessary clauses:
\begin{displaymath}
  \xymatrix@C=1.4em@R=1ex {
    && \\
    p \ar[r] \ar[rd] & {\whardness(F)} \ar[r] \ar[dr] &{\semspace(F)} \aru[r]^{\sim * 3} & {\resspace(F)} \ar[r] & {\treespace(F)} \aru^{=-1}[r] & {\hardness(F)} \ar[r] & \dep(F) \ar[r] & n\\
    & q \ar[r] & {\wid(F)} \ar[urrrr]
  }
\end{displaymath}
Here an arrow ``$h(F) \ra h'(F)$'' means $h(F) \le h'(F)$, and furthermore there exists a sequence $(F_n)$ of clause-sets with bounded $h(F_n)$ but unbounded $h'(F_n)$, while in case of an undirected edge no such separation is possible. The separation $\whardness \ra \semspace$ is shown in \cite{NordstroemHastad2013SpaceLength}, $\resspace \ra \treespace$ in \cite{JaervisaloNordstroem2012PracticalHardness}, and the separation between $\dep$ and $n$ uses unsatisfiable clause-sets which are not minimally unsatisfiable.

\subsection{Extension to satisfiable clause-sets}

These measures do not just apply to unsatisfiable clause-sets, but are extended to \textbf{satisfiable clause-sets}, taking a worst-case approach over all unsatisfiable sub-instances obtained by applying partial assignments (instantiations), or, equivalently, the maximal complexity to derive any (prime) implicate.

\subsubsection{Oblivious polytime SAT solving}

For a fixed bound these measures allow for polynomial-time SAT solving via ``oblivious'' SAT algorithms --- certain basic steps, applied in an arbitrary manner, are guaranteed to succeed. The sets $\Urefc_k$ of all clause-sets $F$ with $\hardness(F) \le k$ yield the basic hierarchy, and we have SAT decision in time $O(n(F)^{2 \hardness(F)-2} \cdot \ell(F))$. The special case $\Urefc_1 = \Urefc$ was introduced in \cite{Val1994UnitResolutionComplete} for the purpose of Knowledge Compilation (KC), and in \cite{GwynneKullmann2012SlurSOFSEM,GwynneKullmann2012SlurJ} we showed that $\Urefc = \Slur$ holds, where $\Slur$ is the class introduced in \cite{SAFS95} as an umbrella class for polynomial-time SAT solving. By \cite{BBCGKV2013Propc,GwynneKullmann2012SlurJ} we get that membership decision for $\Urefc_k$ with $k \ge 1$ is coNP-complete.

\subsubsection{Good representation of boolean functions}

Perhaps the main aim of measuring the complexity of satisfiable clause-sets is to obtain \emph{SAT representations of boolean functions} of various quality (``hardness'') and sizes; see \cite{GwynneKullmann2013GoodRepresentationsII,GwynneKullmann2013GoodRepresentationsIILata} for investigations into XOR-constraints. The motivation is, that we are looking for a ``good'' representation $F$ of a boolean function (like a cardinality or an XOR-constraint) in the context of a larger SAT problem representation. ``Good'' means not ``too big'' and of ``good'' inference power. The latter means (at least) that all unsatisfiable instantiations of $F$ should be easy for SAT solvers, and thus the worst-case approach. In the diagram above, having low $\dep(F)$ is the strongest condition, having low $\whardness(F)$ the weakest. The KC aspects, showing size-hardness trade-offs, are further investigated in \cite{GwynneKullmann2013GoodRepresentationsI}; see Corollary \ref{cor:hierspace2} for an application. This study of the ``best''choice of a representation, considering size (number of clauses) and hardness (like $\hardness$, $\whardness$ or $\semspace$) among all (logically) equivalent clause-sets (at least), likely could not be carried out using (symmetric) width (the current standard), but requires \emph{asymmetric width}, to handle unbounded clause length. The traditional method of reducing the clause-length, by breaking up clauses via auxiliary variables, introduces unnecessary complexity, and can hardly be applied if we only want to consider (logically) equivalent clause-sets (without auxiliary variables).

\subsection{Overview on results}
\label{sec:overviewres}

The theorems (main results):
\begin{enumerate}
\item Theorem \ref{thm:relssts}: relation between resolution-space and semantic-space.
\item Theorem \ref{thm:hd=PI}: new game characterisation of hardness.
\item Theorem \ref{thm:methodwhdlb}: characterisation of w-hardness via partial assignments.
\item Theorem \ref{thm:whdspiel}: game characterisation of w-hardness.
\item Theorem \ref{thm:resspace}: w-hardness is lower bound for semantic space.
\item Theorem \ref{thm:phpwhardness}: hardness and w-hardness for general PHP.
\item Theorem \ref{thm:hdephp}: hardness for EPHP.
\item Theorems \ref{thm:repphp}, \ref{thm:repphpgen}: no polysize representation of bijective PHP.
\item Theorem \ref{thm:2xor}: hardness of two xor-clauses which are together unsatisfiable.
\end{enumerate}

Some references to results of this report are made in \cite{GwynneKullmann2013GoodRepresentationsII} (regarding xor-principles and EPHP), and in the earlier (dis-continued) report \cite{GwynneKullmann2013GoodRepresentations} (regarding PHP).

Chapters \ref{sec:intro} -- \ref{sec:widsemsp} of this report are in a reasonable shape, only the Questions are of a preliminary nature. But the application Chapters \ref{sec:blcl} -- \ref{sec:expxor} are still very preliminary (proofs often not given, and not much explanations).

\section{Preliminaries}
\label{sec:Preliminaries}

We use the general concepts and notations as outlined in \cite{Kullmann2007HandbuchMU}.

\subsection{Clause-sets}
\label{sec:prelimcls}

$\Va$ is the (infinite) set of variables, while $\Lit$ is the set of literals, where every literal is either a variable $v$ or a complemented (negated) variable $\ol{v}$. For a set $L \sse \Lit$ of literals we use $\ol{L} := \set{\ol{x} : x \in L}$. A clause is a finite $C \subset \Lit$ with $C \cap \ol{C} = \es$, the set of all clauses is $\Cl$. A clause-set is a finite set of clauses, the set of all clause-sets is $\Cls$. For $k \in \NNZ$ we define $\Pcls{k}$ as the set of all $F \in \Cls$ where every clause $C \in F$ has length (width) at most $k$, i.e., $\abs{C} \le k$. Via $\var: \Lit \ra \Va$ we assign to every literal its underlying variable; this is extended to clauses $C$ via $\var(C) := \set{\var(x) : x \in \Lit}$, and to clause-sets $F$ via $\var(F) := \bc_{C \in F} \var(C)$. Furthermore we use $\lit(F) := \var(F) \cup \ol{\var(F)}$ for the set of all possible literals over the variables in $F$. The literals occurring in $F \in \Cls$ are given by $\bc F \subset \Lit$. A literal $x$ is called \emph{pure for $F$} if $\ol{x} \notin \bc F$.

Measures for $F \in \Cls$ are $n(F) := \abs{\var(F)} \in \NNZ$ (number of variables), $c(F) := \abs{F} \in \NNZ$ (number of clauses), and $\ell(F) := \sum_{C \in F} \abs{C} \in \NNZ$ (number of literal occurrences). A special clause is the empty clause $\bot := \es \in \Cl$, a special clause-set is the empty clause-set $\top := \es \in \Cls$.

A partial assignment is a map $\vp: V \ra \set{0,1}$ for some finite $V \subset \Va$, the set of all partial assignments is $\Pass$. The number of variables in a partial assignment is denoted by $n(\vp) := \abs{\var(\vp)}$. For a clause $C$ we denote by $\vp_C \in \Pass$ the partial assignment which sets precisely the literals in $C$ to $0$. The application (instantiation) of $\vp$ to $F \in \Cls$ is denoted by $\vp * F \in \Cls$, obtained by first removing satisfied clauses $C \in F$ (i.e., containing a literal $x \in C$ with $\vp(x) = 1$), and then removing all falsified literals from the remaining clauses.

The set of satisfiable clause-sets is $\Sat := \set{F \in \Cls \mb \ex\, \vp \in \Pass : \vp * F = \top}$, while $\Usat := \Cls \sm \Sat$ is the set of unsatisfiable clause-sets.

For $F, F' \in \Cls$ the implication-relation is defined as usual: $F \models F' :\Lra \fa\, \vp \in \Pass : \vp * F = \top \Ra \vp * F' = \top$. We write $F \models C$ for $F \models \set{C}$. A clause $C$ with $F \models C$ is an \emph{implicate} of $F$, while a \emph{prime implicate} is an implicate $C$ such that no $C' \subset C$ is also an implicate; $\primec_0(F)$ is the set of prime implicates of $F$.

\subsection{Resolution}
\label{sec:prelimres}

\begin{defi}\label{def:resolution}
  Two clauses $C, D$ are \textbf{resolvable} if $\abs{C \cap \ol{D}} = 1$, i.e., they clash in exactly one variable:
  \begin{itemize}
  \item For two resolvable clauses $C$ and $D$ the \textbf{resolvent} $\bmm{C \res D} := (C \cup D) \sm \set{x,\ol{x}}$ for $C \cap \ol{D} = \set{x}$ is the union of the two clauses minus the resolution literals.
  \item $x$ is called the \textbf{resolution literal}, while $\var(x)$ is the \textbf{resolution variable}.
  \end{itemize}
\end{defi}
Remarks:
\begin{enumerate}
\item If $x$ is the resolution variable of $C, D$, then $x \in C$, and $\ol{x}$ is the resolution variable of $D, C$.
\item The closure of $F \in \Cls$ under resolution is a clause-set with $\primec_0(F)$ as its subsumption-minimal elements.
\end{enumerate}

The set of nodes of a tree $T$ is denoted by $\nds(T)$, the set of leaves by $\lvs(T) \sse \nds(T)$. The height $\height_T(w) \in \NNZ$ of a node $w \in \nds(T)$ is the height of the subtree of $T$ rooted at $w$ (so $\lvs(T) = \set{w \in \nds(T) : \height_T(w) = 0}$).
\begin{defi}\label{def:resproof}
  A \textbf{resolution tree} is a pair $R = (T,C)$ such that:
  \begin{itemize}
  \item $T$ is an ordered rooted tree, where every inner node has exactly two children, and where the set of nodes is denoted by $\nds(T)$ and the root by $\rt(T) \in \nds(T)$.
  \item While $C: \nds(T) \ra \Cls$ labels every node with a clause such that the label of an inner node is the resolvent of the labels of its two parents.
  \end{itemize}
  We use:
  \begin{itemize}
  \item $\bmm{\premr(R)} := \set{\concr(w) : w \in \lvs(T)} \in \Cls$ for the ``axioms'' (or ``premisses'') of $R$;
  \item $\bmm{\concr(R)} := C(\rt(T)) \in \Cl$ as the ``conclusion'';
  \item $\bmm{\allcr(R)} := \set{C(w) : w \in \nds(T)} \in \Cls$ for the set of all clauses in $R$.
  \end{itemize}
  A \textbf{resolution proof} $R$ of a clause $C$ from a clause-set $F$, denoted by \bmm{R: F \vdash C}, is a resolution tree $R = (T,C)$ such that
  \begin{itemize}
  \item $\premr(R) \sse F$,
  \item $\concr(R) = C$.
  \end{itemize}
  We use \bmm{F \vdash C} if there exists a resolution proof $R$ of some $C' \sse C$ from $F$ (i.e., $R: F \vdash C'$). A \textbf{resolution refutation} of a clause-set $F$ is a resolution proof deriving $\bot$ from $F$. 
  \begin{itemize}
  \item The \textbf{tree-resolution complexity} $\bmm{\comptr(R)} \in \NN$ is the number of leaves in $R$, that is, $\comptr(R) := \nlvs(R) = \abs{\lvs(T)}$.
  \item The \textbf{resolution complexity} $\bmm{\compr(R)} \in \NN$ is the number of \emph{distinct} clauses in $R$, that is $\compr(R) := c(\allcr(R))$.
  \end{itemize}
  Finally, for $F \in \Usat$ we set
  \begin{itemize}
  \item $\bmm{\comptr(F)} := \min \set{\comptr(R) \mb R : F \vdash \bot} \in \NN$
  \item $\bmm{\compr(F)} := \min \set{\compr(R) \mb R : F \vdash \bot} \in \NN$.
  \end{itemize}

\end{defi}
Typically we identify $R = (T,C)$ with $T$, while suppressing the labelling $C$. Note that we use resolution \emph{trees} also when speaking about full resolution, which enables us to use branching/splitting trees also for the analysis of full resolution, at least in our context; see Subsection \ref{sec:prelimtreesres} for more on that.

\subsection{Extension of measures to satisfiable clause-sets}
\label{sec:extmeasures}

Before we start to define individual hardness measures, we introduce our general method for extending measures for unsatisfiable clause-sets to arbitrary clause-sets, both unsatisfiable and satisfiable. This is quite important as --- in sharp contrast to the situation for unsatisfiable formulas --- very little is known from the theoretical side about the complexity of SAT solvers on satisfiable instances. The special case of extension of hardness to satisfiable clause-sets was first mentioned by Ans\'{o}tegui et al.\ \cite{AnsoteguiBonetLevyManya2008Hardness}, and introduced and investigated, in a more general form, in \cite{GwynneKullmann2012SlurJ,GwynneKullmann2013GoodRepresentationsIILata}.

Every measure $h_0: \Usat \ra \NNZ$ with the property $\fa\, F \in \Usat\, \fa\, \vp \in \Pass : h_0(\vp * F) \le h_0(F)$ is extended to $h: \Cls \ra \NNZ$ by
\begin{enumerate}
\item $h(\top) := \min_{F \in \Usat} h_0(F)$.
\item For $F \in \Cls \sm \set{\top}$ we define $h(F)$ as the maximum of $h_0(\vp * F)$ for $\vp \in \Pass$ with $\vp * F \in \Usat$.
\end{enumerate}
So we get $\fa\, F \in \Cls\, \fa\, \vp \in \Pass : h(\vp * F) \le h(F)$, and $h(F) = h_0(F)$ for $F \in \Usat$. And for $h_0 \le h_0'$ we get $h \le h'$. Note that for the computation of $h(F)$ as the maximum of $h_0(\vp * F)$ for unsatisfiable $\vp * F$, one only needs to consider minimal $\vp$ (since application of partial assignments can not increase the measure), that is, $\vp_C$ for $C \in \primec_0(F)$; so for $F \in \Cls \sm \set{\top}$ we have $h(F) = \max_{C \in \primec_0(F)} h_0(\vp * F)$. For $V \sse \Va$ the relativised version $\bmm{h^V}: \Cls \ra \NNZ$ is defined by only considering partial assignments $\vp \in \Pass$ with $\var(\vp) \sse V$.

In the following we will therefore define the hardness measure only for unsatisfiable clauses and then extend them via the above method.

\subsection{Tree-hardness}
\label{sec:prelimhd}

We start with what is in our opinion one of the central hardness measures for resolution, which is why we simply call it \emph{hardness} (but for differentiation it might be called \emph{tree-hardness}, then written ``$\thardness$'').
It seems that this concept was reinvented in the literature several times. Intuitively, the hardness measures the height of the biggest full binary tree which can be embedded into each tree-like resolution refutation of the formula. This is also known as the Horton-Strahler number of a tree (see \cite{Viennot1990Trees,EsparzaLuttenbergerSchlund2014Strahler}). In the context of resolution this measure was first introduced in \cite{Ku99b,Ku00g}, and extended in \cite{GwynneKullmann2012SlurSOFSEM,GwynneKullmann2012SlurJ}:
\begin{defi}\label{def:hd}
  For $F \in \Usat$ let $\bmm{\hardness(F)} \in \NNZ$ be the minimal $k \in \NNZ$ such that a resolution tree $T: F \vdash \bot$ exists, where the Horton-Strahler number of $T$ is at most $k$, that is, for every node in $T$ there exists a path to some leaf of length at most $k$. For $k \in \NNZ$ let $\bmm{\Urefc_k} := \set{F \in \Cls : \hardness(F) \le k}$.
\end{defi}

See \cite{Ku99b,Ku00g,GwynneKullmann2012SlurSOFSEM,GwynneKullmann2012SlurJ} for the various equivalent description, where especially the algorithmic approach, via generalised unit-clause propagation $\rk_k$, is notable: hardness is the minimal level $k$ of generalised unit-clause propagation needed to derive a contradiction under any instantiation. As shown in \cite[Corollary 7.9]{Ku99b}, and more generally in \cite[Theorem 5.14]{Ku00g}, we have
\begin{displaymath}
  2^{\hardness(F)} \le \comptr(F) \le (n(F)+1)^{\hardness(F)}
\end{displaymath}
for $F \in \Usat$.
A simpler measure is the minimum depth of resolution refutations:
\begin{defi}\label{def:dep}
  For $F \in \Usat$ let $\bmm{\dep(F)} \in \NNZ$ be the minimal height of a resolution tree $T: F \vdash \bot$.
\end{defi}

Remarks:
\begin{enumerate}
\item Since the Horton-Strahler number of a tree is at most the height, we get $\hardness(F) \le \dep(F)$ for all $F \in \Cls$.
\item For $k \in \NNZ$ the class of $F \in \Cls$ with $\dep(F) \le k$ is called $\Canoni{k}$ in \cite{CepekKuceraVlcek2012SLUR,BalyoGurskyKuceraVlcek2012SLURHier}. Obviously $\Canoni{0} = \Urefc_0$. See Subsection 7.2 in \cite{GwynneKullmann2012SlurJ} and Subsection 9.2 in \cite{GwynneKullmann2013GoodRepresentationsI} for further results.
\item See Subsection \ref{sec:characdep} for more results.
\end{enumerate}

\subsection{Width-hardness}
\label{sec:prelimwhd}

The standard resolution-width of an unsatisfiable clause-set $F$ is the minimal $k$ such that a resolution refutation of $F$ using only clauses of length at most $k$ exists:
\begin{defi}\label{def:swid}
  For $F \in \Usat$ the \textbf{symmetric width} $\bmm{\wid(F)} \in \NNZ$ is the smallest $k \in \NNZ$ such that there is $T: F \vdash \bot$ with $\allcr(T) \in \Pcls{k}$.
\end{defi}

Based on the notion of ``$k$-resolution'' introduced in \cite{Kl93}, the ''asymmetric width'' was introduced in \cite{Ku99b,Ku000,Ku00g} (and further studied in \cite{GwynneKullmann2012SlurJ,GwynneKullmann2013GoodRepresentationsI,GwynneKullmann2013GoodRepresentationsIILata}). Different from the symmetric width, only \emph{one parent clause} needs to have size at most $k$ (while there is no restriction on the other parent clause nor on the resolvent):
\begin{defi}\label{def:whd}
  For a resolution tree $T$ its \textbf{(asymmetric) width} $\bmm{\whardness(T)} \in \NNZ$ is defined as $0$ if $T$ is trivial (i.e., $\abs{\nds(T)} = 1$), while otherwise for left and right children $w_1, w_2$ with subtrees $T_1, T_2$ we define
  \begin{displaymath}
    \whardness(T) := \max \big (\whardness(T_1), \whardness(T_2), \, \min(\abs{C(w_1)}, \abs{C(w_2)}) \big )
  \end{displaymath}
  (note that the corresponding definition of $\wid(T)$ just has the $\min$ replaced by a (second) $\max$). We write \bmm{R : F \vdash^k C} if $R : F \vdash C$ and $\whardness(R) \le k$. Now for $F \in \Usat$ we define $\bmm{\whardness(F)} := \min \set{\whardness(T) \mb T : F \vdash \bot}\in \NNZ$. For $k \in \NNZ$ let $\bmm{\Wrefc_k} := \set{F \in \Cls : \whardness(F) \le k}$.
\end{defi}
Basic properties of w-hardness are:
\begin{enumerate}
\item $\Wrefc_0 = \Urefc_0$ and $\Wrefc_1 = \Urefc_1$.
\item For all $F \in \Cls$ holds $\whardness(F) \le \hardness(F)$ (for unsatisfiable $F$ this is shown in Lemma 6.8 in \cite{Ku00g}, which extends to satisfiable clause-sets by definition; in Lemma \ref{lem:whdhdg} we provide a new proof).
\item For the relation between $\wid$ and $\whardness$ see Subsection \ref{sec:relsymmasymw}.
\item In \cite{SW98J} a fundamental relation between symmetric width and proof size for resolution refutations is observed, thereby establishing one of the main methods to prove resolution lower bounds. In \cite[Theorem 8.11]{Ku99b} and \cite[Theorem 6.12, Lemma 6.15]{Ku00g}
this size-width relation is strengthened to asymmetric width:
\begin{displaymath}
  e^{\frac 18 \frac{\whardness(F)^2}{n(F)}} < \compr(F) < 6 \cdot n(F)^{\whardness(F)+2}
\end{displaymath}
for $F \in \Usat \sm \set{\set{\bot}}$, where $e^{\frac 18} = 1.1331484 \ldots$ Note that compared to \cite{SW98J} the numerator of the exponent does not depend on the maximal clause-length of $F$.
\item In \cite[Lemma 8.13]{Ku99b} it is shown that the partial ordering principle has asymmetric width the square-root of the number of variables, while having a polysize resolution refutation.
\end{enumerate}

\begin{examp}\label{exp:whd}
  Some easy examples for $\wid(F)$ and $\whardness(F)$:
  \begin{enumerate}
  \item $\wid(\set{\bot}) = \whardness(\set{\bot}) = 0$.
  \item More generally for $C \in \Cl$ holds $\wid(\set{C}) = \whardness(\set{C}) = 0$.
  \item In general we have $\wid(F) = 0 \Lra \whardness(F) = 0$ for all $F \in \Cls$.
  \item For $F := \set{\set{a},\set{\ol{a},b},\set{\ol{a},\ol{b}}}$ we have $\wid(F) = 2$ and $\whardness(F) = 1$.
  \item For a Horn clause-set $F$ holds $\whardness(F) \le 1$ (since unit-clause propagation is sufficient to derive unsatisfiability), while $\wid(F)$ is unbounded (if $F$ is minimally unsatisfiable, then $\wid(F)$ equals the maximal clause-length of $F$).
  \item For general minimally unsatisfiable $F$, the maximal clause-length is a lower bound for $\wid(F)$, but is unrelated to $\whardness(F)$. (For bounded clause-length however, $\wid$ and $\whardness$ can be considered asymptotically equivalent by Corollary \ref{cor:inputressim}.)
  \end{enumerate}
\end{examp}

\subsection{Trees and (full!) resolution}
\label{sec:prelimtreesres}

As it is widely known, and where more details can be found in \cite{Ku99b,Ku00g}, resolution trees are closely related to ``splitting'' or ``branching'' trees, full binary trees labelled with clause-sets and corresponding to the backtracking tree of the simplest recursive SAT solver on unsatisfiable inputs. In Theorem 7.5 in \cite{Ku99b} and in a more general form in Subsection 5.2 in \cite{Ku00g} the close relation between branching trees and (regular!) resolution trees is discussed. Now it appears that this connection breaks when it comes to full resolution, but this is actually only partially so: regarding the number of different clauses in a resolution tree, it is known that regularisation can indeed exponentially increase the number of different clauses, however when it comes to width, symmetric or asymmetric, then there are no difficulties, since the process of regularisation, implicit in the correspondences between resolution trees and branching trees, does never increase clause-sizes. So for the resolution trees used in Definitions \ref{def:swid}, \ref{def:whd} of symmetric resp.\ asymmetric width, w.l.o.g.\ one can restrict attention to regular resolution trees or resolution trees derived from branching trees.

Formally, the branching trees for a clause-set $F \in \Usat$ are the full binary trees obtained as follows:
\begin{enumerate}
\item If $\bot \in F$, then then only branching tree for $F$ is the one-node tree labelled with $F$.
\item Otherwise, the branching trees for $F$ are obtained by choosing a variable $v \in \Va$, labelling the root with $F$, and choosing a branching tree for $\pao v0 * F$ as left subtree and a branching tree for $\pao v1 * F$ as right subtree.
\end{enumerate}
For a node $w \in \nds(T)$ of a branching tree, we denote the partial assignment collecting the assignments along the edges from the root to $w$ by $\bmm{\vp(w)} \in \Pass$. We call a branching tree $T$ \emph{minimal}, if it is obtained back without change after first translating $T$ into a resolution refutation for $F$, and then translating this refutation back into a branching tree for $F$. In other words, no branching in a minimal branching tree is superfluous.

\section{Space complexity}
\label{sec:defspacecomp}

The last measures that we discuss in this paper relate to space complexity. We consider three measures: \emph{semantic space}, \emph{resolution space} and \emph{tree space}.

\subsection{Semantic space}
\label{sec:spacesem}

Semantic space was introduced in \cite{AlekhnovichBRW02}; a slightly modified definition follows.
\begin{defi}\label{def:spaceres}
  Consider $F \in \Cls$ and $k \in \NN$. A \textbf{semantic $k$-sequence for $F$} is a sequence $F_1, \dots, F_p \in \Cls$, $p \in \NN$, fulfilling the following conditions:
  \begin{enumerate}
  \item For all $i \in \tb 1p$ holds $c(F_i) \le k$.
  \item $F_1 = \top$, and for $i \in \tb 2p$ either holds
    \begin{enumerate}
    \item $F_{i-1} \models F_i$ (inference), or
    \item there is $C \in F$ with $F_i = F_{i-1} \cup \set{C}$ (axiom download).
    \end{enumerate}
  \end{enumerate}
  A semantic sequence is called \textbf{complete} if $F_p \in \Usat$. For $F \in \Usat$ the \textbf{semantic-space complexity} of $F$, denoted by $\bmm{\semspace(F)} \in \NN$, is the minimal $k \in \NN$ such there is a complete semantic $k$-sequence for $F$.
\end{defi}
Remarks:
\begin{enumerate}
\item $\semspace(F) = 1$ iff $\bot \in F$.
\item Every non-empty initial part of a semantic $k$-sequence for $F$ is also a semantic $k$-sequence for $F$.
\item We have $F \models F_i$ for $i \in \tb 1p$ and a semantic sequence $F_1,\dots,F_p$ for $F$.
\item Different from \cite{AlekhnovichBRW02}, the elimination of clauses (``memory erasure'') is integrated into the inference step, since we want our bound $\whardness \le \semspace$ (Theorem \ref{thm:resspace}) to be as tight as possible, and the tree-space, as a special case of semantic space, shall fulfil $\treespace = \hardness + 1$ (Lemma \ref{lem:spacetres}).
\item One could integrate both possibilities (inference and download), also allowing several downloads at once, into the single condition ``$F_{i-1} \models F_i \sm F$'' (similar to \cite{To99,ET99b}, there for resolution space). However, different from the integration of the removal of clauses into the inference step, this does not lead to smaller space, since we can first proceed $F_{i-1} \leadsto F_i \sm F$, and then adding one axiom after another. Requiring the addition of axioms as a separate step, and one by one, seems the most useful formulation for proofs.
\end{enumerate}

\begin{examp}\label{exp:semspaceeqwhard}
  Consider
  \begin{multline*}
    F := \setb { \set{1,2,3}, \set{-1,2,3}, \set{1,-2,3}, \set{-1,-2,3},\\
      \set{4,5,6}, \set{-4,5,6}, \set{4,-5,6},\set{-4,-5,6},\\
      \set{-3,-6} }.
  \end{multline*}
  The following is a semantic 4-sequence for $F$ (where for convenience we compress several axiom downloads into one step):
  \begin{enumerate}
  \item $F_1 := \set{\set{1,2,3}, \set{-1,2,3}}$.
  \item $F_2 := \set{\set{2,3}}$.
  \item $F_3 := \set{\set{2,3}, \set{1,-2,3}, \set{-1,-2,3}}$.
  \item $F_4 := \set{\set{3}}$.
  \item $F_5 := \set{\set{3}, \set{4,5,6}, \set{-4,5,6}}$.
  \item $F_6 := \set{\set{3}, \set{5,6}}$.
  \item $F_7 := \set{\set{3}, \set{5,6}, \set{4,-5,6},\set{-4,-5,6}}$.
  \item $F_8 := \set{\set{3}, \set{6}}$.
  \item $F_9 := \set{\set{3}, \set{6}, \set{-3,-6}}$.
  \end{enumerate}
\end{examp}

\begin{examp}\label{exp:semspaceHO}
  $\semspace(\set{\set{v}}) = \semspace(\set{\set{\ol{v},\ol{w}}}) = 1$ and $\semspace(\set{\set{v},\set{\ol{v}}}) = 2$. More generally, for Horn clause-sets $F \in \Ho$ we have:
  \begin{enumerate}
  \item $\semspace(F) \le 2$.
  \item If $F \in \Usat$: $\semspace(F) = 2$ iff $\bot \notin F$.
  \end{enumerate}
\end{examp}

\begin{lem}\label{lem:rksemspace}
  For $F \in \Usat$ and $\vp \in \Pass$ holds $\semspace(\vp * F) \le \semspace(F)$.
\end{lem}
\begin{prf}
We note that for arbitrary $F, G \in \Cls$ and $\vp \in \Pass$ we have $F \models G \Ra \vp * F \models \vp * G$. Thus, if $(F_1,\dots,F_p)$ is a complete semantic sequence for $F$, then $(\vp * F_1,\dots,\vp * F_p)$ is a complete semantic sequence for $\vp * F$. \Qed
\end{prf}

\begin{lem}\label{lem:standsemseq}
  If there is a semantic $k$-sequence $F_1,\dots,F_p$ for $F \in \Cls$ and $k \in \NN$, then there is a semantic $k$-sequence $F'_1,\dots,F'_q$ for $F$ with $q \le p$ such that:
  \begin{enumerate}
  \item $\var(F'_i) \sse \var(F)$ for $i \in \tb 1q$.
  \item For all $i \in \tb 1q$ the clause-set $F_i'$ is irredundant.
  \item An inference step is not followed by another inference step.
  \item For an axiom download $F_{i+1}' = F_i' \cup \set{C}$ we have $C \notin F_i'$ (i.e., $c(F_{i+1}') > c(F_i')$).
  \item For an inference step $F_i' \models F_{i+1}'$ we have $F_{i+1}' \sse \primec_0(F_i')$. Moreover, $F_{i+1}'$ has the minimal number of clauses amongst all clause-sets equivalent to $F_{i+1}'$, and we have $c(F_{i+1}') < c(F_i')$.
  \item For all $i \in \tb 1p$ there is $j \in \tb 1q$ with $F'_j \models F_i$.
  \item For $i \in \tb{1}{q-1}$ holds $F'_i \in \Sat$.
  \end{enumerate}
  We call a sequence fulfilling these conditions \textbf{standardised}.
\end{lem}
\begin{prf}
In general holds, that in a semantic sequence $F_1,\dots,F_p$ for $F$, for every inference step $F_i \leadsto F_{i+1}$ the clause-set $F_{i+1}$ can be replaced by any clause-set $F'$ with $F_i \models F' \models F_{i+1}$, and we still have a semantic sequence for $F$. And for two successive inference steps, the first step can be left out. If at some point an unsatisfiable clause-set is obtained, then a possible remainder of the sequence can be cut off. Via these operations we obtain a desired standardisation. \Qed
\end{prf}

Remarks:
\begin{enumerate}
\item In a standardised semantic sequence $F_1,\dots,F_p$ for an axiom download $F_{i+1} = F_i \cup \set{C}$ we have $F_i \not\models C$, and if it is complete, then $F_p \in \Musat$, and the last step $F_{p-1} \leadsto F_p$ is an axiom download.
\item A standardised semantic sequence consists of two alternating actions:
  \begin{enumerate}
  \item An \emph{expansion step}, a sequence of axiom downloads, where each added axiom is new, and the finally obtained clause-set is irredundant.
  \item A \emph{contraction step}, which takes the set $\primec_0(F_i)$ of prime implicates of the current $F_i$, and selects a subset $F_{i+1} \subset \primec_0(F_i)$ with $c(F_{i+1}) < c(F_i)$, such that $F_{i+1}$ is not only irredundant, but also there is no clause-set $F'$ equivalent to $F_{i+1}$ with $c(F') < c(F_{i+1})$.
  \end{enumerate}
\end{enumerate}

\begin{lem}\label{lem:semspacemunf}
  Consider $F \in \Usat$ and a complete semantic $k$-sequence $F_1,\dots,F_p$ ($k \in \NN$) for $F$. Then there is a standardised complete semantic $k$-sequence $F_1',\dots,F_p'$ for $F$ and $C \in F$ with $F_{p-1}' = \set{\set{\ol{x}} : x \in C}$ and $F_p' = F_{p-1}' \cup \set{C}$ (thus $\abs{C} \le k-1$).
\end{lem}
\begin{prf}
By the standardisation condition the last step is an axiom download: $F_p = F_{p-1} \cup \set{C}$ for some $C \in F$. We have $F_p \in \Musat$, and so for every $x \in C$ we have $F_{p-1} \models \set{\ol{x}}$. If $F_{p-1}$ was obtained by an inference step, then we have already $F_{p-1} = \set{\set{\ol{x}} : x \in C}$, while otherwise we insert one inference step. \Qed
\end{prf}

\begin{corol}\label{cor:semspacegek}
  Consider $k \in \NNZ$ and $F \in \Usat$ with $\fa\, C \in F : \abs{C} \ge k$. Then $\semspace(F) \ge k+1$.
\end{corol}

\begin{lem}\label{lem:semspspc}
  For $F \in \Usat$ holds:
  \begin{enumerate}
  \item $\whardness(F) = 0 \Lra \semspace(F) = 1$.
  \item $\whardness(F) = 1 \Lra \semspace(F) = 2$.
  \item $\whardness(F) \ge 2 \Ra \semspace(F) \ge 3$.
  \end{enumerate}
\end{lem}
\begin{prf}
We have $\whardness(F) = 0$ iff $F = \set{\bot}$ iff $\semspace(F) = 1$. If $\whardness(F) = 1$, then $\hardness(F) = 1$, and thus $\semspace(F) \le 2$, whence by the first part $\semspace(F) = 2$. Finally assume $\whardness(F) \ge 2$, and we have to show $\semspace(F) \ge 3$. Since $\semspace(\rk_1(F)) \le \semspace(F)$ (Lemma \ref{lem:rksemspace}), and furthermore $\whardness(F) = \whardness(\rk_1(F))$ (due to $\whardness(F) \ge 2$), we can assume $\rk_1(F) = F$. By the first part we have $\semspace(F) \ge 2$. Assume $\semspace(F) = 2$, and consider a semantic $2$-sequence $(F_1,\dots,F_p)$. We have $F_p \in \Musat$ with $c(F_p) = 2$, and thus there is a variable $v$ with $F_p$ isomorphic $\set{\set{v},\set{\ol{v}}}$. Since $F_p$ is obtained by axiom download, we have a contradiction to the assumption, that $F$ does not contain unit-clauses. \Qed
\end{prf}

\begin{corol}\label{cor:semspspc}
  For $F \in \Pcls{2}$ with $\whardness(F) = 2$ we have $\semspace(F) = 3$.
\end{corol}
\begin{prf}
For $F \in \Pcls{2}$ we have $\hardness(F) \le 2$ (see \cite{GwynneKullmann2012SlurJ}), and thus $\semspace(F) \le 3$. \Qed
\end{prf}

\subsection{Resolution space and tree space}
\label{sec:resspace}

We come to the notion of resolution space originating in \cite{KlLe94,KlLe99} and \cite{To99,ET99b}. This measure was intensively studied during the last decade (cf.\ e.g.\ \cite{BN11,Nordstrom2012PebbleProofTimeSpaceSurvey}).
\begin{defi}\label{def:spaceresprop}
  Consider $F \in \Cls$ and $k \in \NN$. A \textbf{resolution $k$-sequence for $F$} is a sequence $F_1, \dots, F_p \in \Cls$, $p \in \NN$, fulfilling the following conditions:
  \begin{enumerate}
  \item For all $i \in \tb 1p$ holds $c(F_i) \le k$.
  \item $F_1 = \top$, and for $i \in \tb 2p$ either holds
    \begin{enumerate}
    \item $F_i \sm F_{i-1} = \set{C}$, where $C$ is a resolvent of two clauses in $F_i$ (removal of clauses and/or addition of one resolvent), or
    \item there is $C \in F$ with $F_i = F_{i-1} \cup \set{C}$ (axiom download).
    \end{enumerate}
  \end{enumerate}
  A resolution $k$-sequence is \textbf{complete} if $\bot \in F_p$. For $F \in \Usat$ the \textbf{resolution-space complexity} of $F$, denoted by $\bmm{\resspace(F)} \in \NN$, is the minimal $k \in \NN$ such there is a complete resolution $k$-sequence for $F$.
\end{defi}
Thus a (complete) resolution $k$-sequence for $F$ is a (complete) semantic $k$-sequence for $F$:
\begin{lem}\label{lem:relssrs}
  For $F \in \Cls$ holds $\semspace(F) \le \resspace(F)$.
\end{lem}
An alternative definition of $\resspace(F)$ for $F \in \Usat$ uses resolution dags and ``pebbling games'' (\href{http://en.wikipedia.org/wiki/Pebble_game}{Black Pebbling}): On a dag $G$ with a unique sink the pebbling game allows to place pebbles on sources, and if all direct predecessors of a node are pebbled, then also that node can be pebbled, where the pebbles on the predecessors can be kept or they can be deleted (individually, at the same time when the node is pebbled). The pebbling number $\peb(G) \in \NN$ is the minimum number of pebbles needed to pebble the sink. Now it is easy to see that $\resspace(F) = \peb(F)$, where $\peb(F)$ is the minimum of $\peb(G)$ for resolution-dags $G$ refuting $F$.

\begin{defi}\label{def:treeres}
  A \textbf{tree $k$-sequence for $F$} is a resolution $k$-sequence for $F$, such that in case of adding an inferred clause via $F_i \sm F_{i-1} = \set{R}$, for $R = C \res D$ with $C, D \in F_{i-1}$, we always have $C, D \notin F_i$. For $F \in \Usat$ the \textbf{tree-resolution space complexity} of $F$, denoted by $\bmm{\treespace(F)} \in \NN$, is the minimal $k \in \NN$ such there is a complete tree $k$-sequence for $F$.
\end{defi}

By definition we have:
\begin{lem}\label{lem:relrsts}
  For $F \in \Cls$ holds $\resspace(F) \le \treespace(F)$.
\end{lem}

\begin{lem}\label{lem:resspspc}
  For $F \in \Usat$ holds:
  \begin{enumerate}
  \item\label{lem:resspspc1} $\semspace(F) = 1 \Lra \resspace(F) = 1 \Lra \treespace(F) = 1 \Lra \hardness(F) = 0 \Lra \bot \in F$.
  \item\label{lem:resspspc2} $\semspace(F) = 2 \Lra \resspace(F) = 2 \Lra \treespace(F) = 2 \Lra \hardness(F) = 1 \Lra \rk_1(F) = \set{\bot}$.
  \end{enumerate}
\end{lem}
\begin{prf}
Part \ref{lem:resspspc1} is trivial. For Part \ref{lem:resspspc2} we have $\semspace(F) = 2 \Lra \whardness(F) = 1 \Lra \hardness(F) = 1 \Lra \treespace(F) = 2$. \Qed
\end{prf}

\begin{conj}\label{con:ss3}
  For $F \in \Usat$ holds $\semspace(F) = 3 \Ra \resspace(F) = 3$.
  \begin{enumerate}
  \item By considering what can happen with the inference step, this shouldn't be too hard to prove.
  \item See Question \ref{que:relssts} for the general context.
  \item If $\semspace(F) = 3$, then $\whardness(F) \in \set{2,3}$. Can the case $\whardness(F) = 3$ be excluded here? (The answer would be yes, if Conjecture \ref{con:improvesslb} would hold.)
  \item Compare Question \ref{que:whd2ss} (there we ask for more complicated (unsatisfiable) clause-sets, with $\whardness(F) = 2$ and unbounded $\semspace(F)$).
  \end{enumerate}
\end{conj}

As shown in Subsection 7.2.1 in \cite{Ku99b}, for $F \in \Usat$ we have $\treespace(F) = \hardness(F) + 1$. The proof uses the characterisation of resolution-space via the above Black-Pebbling game: For trees $G$ there is no point in keeping pebbles on predecessors of the node just pebbled, and we see that $\treespace(F)$ equals the minimum of $\peb(T)$ for $T: F \vdash \bot$. Now it is well-known that for binary trees $T$ holds $\hts(T) = \peb(T) + 1$ (where $\hts(T)$ is the Horton-Strahler number of $T$; recall Definition \ref{def:hd}), see for example \cite{Viennot1990Trees,EsparzaLuttenbergerSchlund2014Strahler}.
\begin{lem}\label{lem:spacetres}
  For $F \in \Cls$ holds $\treespace(F) = \hardness(F) + 1$.
\end{lem}

Since $\hardness(F) \le n(F)$:
\begin{corol}\label{cor:spacetres}
  For $F \in \Cls$ holds $\treespace(F) \le n(F) + 1$.
\end{corol}

Since $\hardness(F) \le c(F) - 1$ for unsatisfiable $F$:
\begin{corol}\label{cor:spacetres2}
  For $F \in \Cls$ holds $\treespace(F) \le c(F)$.
\end{corol}

From Corollary \ref{cor:spacetres2} we obtain the more general form:
\begin{corol}\label{cor:spacetres3}
  For $F \in \Cls$ and a prime implicate $C \in \primec_0(F)$ exists a tree $c(F)$-sequence $F_1,\dots,F_p$ for $F$ with $F_p = \set{C}$.
\end{corol}

We remarked earlier that by definition we have $\semspace(F) \le \resspace(F)$. In fact, the two measures are the same up to the factor $2$, as shown by \cite{AlekhnovichBRW02}. This illustrates that space is quite a robust measure, which does not dependent on syntactic details of the resolution calculus. Our factor is $3$, due to the integration of clause-removal and inference. The proof is simpler than in \cite{AlekhnovichBRW02}, not using ``Tarsi's lemma'' (see \cite{Kullmann2007HandbuchMU}).
\begin{thm}\label{thm:relssts}
  For $F \in \Cls$ we have $\resspace(F) \le 3 \semspace(F) - 2$.
\end{thm}
\begin{prf}
Assume $F \in \Usat$. For $k = 1$ the assertion is trivial, so consider $k \ge 2$. Consider a complete semantic $k$-sequence $F_1,\dots,F_p$ for $F$. To obtain a complete resolution $k'$-sequence for $F$ with $k' := 3 k - 2$, we replace every inference step $F_i \leadsto F_{i+1}$ as follows:
\begin{enumerate}
\item We have $c(F_i) \le k$, and w.l.o.g.\ $c(F_{i+1}) \le k-1$.
\item We keep all clauses of $F_i$, until all clauses of $F_{i+1}$ have been derived, as additions to $F_i$.
\item We obtain one clause of $F_{i+1}$ after another, using additionally space at most $c(F_i)-1$ according to Corollary \ref{cor:spacetres3}; note that the clauses of $F_i$ are available, due to keeping $F_i$, and this yields also the reduction ``$-1$''.
\end{enumerate}
So the clause-sets used in such a resolution-sequence have size at most $c(F_i) + c(F_{i+1}) + (c(F_i) - 1) = 3k - 2$. \Qed
\end{prf}

\begin{quest}\label{que:relssts}
  Is the factor $3$ in Theorem \ref{thm:relssts} optimal?
  \begin{enumerate}
  \item Could we even have $\fa\, F \in \Usat : \resspace(F) = \semspace(F)$ ?
  \item I'm not aware of a counter-example. In the light of Conjecture \ref{con:ss3}, the simplest counter-example would have $\semspace(F) = 4$ and $\resspace(F) > 4$ (by Theorem \ref{thm:relssts} we know $\resspace(F) \le 10$).
  \end{enumerate}
\end{quest}

\begin{examp}
  For $n \in \NNZ$ let $A_n$ be the full clause-set over variables $1,\dots,n$ with $2^n$ clauses. By Corollary \ref{cor:semspacegek} we have $\semspace(A_n) \ge n+1$. Since $\hardness(A_n) = n$, we have $\semspace(A_n) = \resspace(A_n) = n+1$. We also have $\whardness(A_n) = n$.
\end{examp}

\begin{quest}\label{que:polytss}
  For fixed $k \in \NNZ$ and input $F \in \Usat$ we can decide in polynomial time whether $\hardness(F) \le k$ or $\whardness(F) \le k$ holds (the former needs only linear space).
  \begin{enumerate}
  \item Can we also decide $\resspace(F) \le k$ in polynomial time?
  \item And what about $\semspace(F) \le k$ ?
  \end{enumerate}

\end{quest}

\section{Tree-hardness}
\label{sec:hdproofct}

The following fundamental lemma shows how the hardness is affected when one variable is assigned to a 0/1 value:
\begin{lem}\label{lem:hdpass}
  For $F \in \Cls$ and $v \in \var(F)$ either
  \begin{enumerate}
  \item\label{lem:hdpass1} there is $\ve \in \set{0,1}$ with $\hardness(\pao v{\ve} * F) = \hardness(F)$ and $\hardness(\pao v{\ol{\ve}} * F) \le \hardness(F)$,
  \item\label{lem:hdpass2} or $\hardness(\pao v0 * F) = \hardness(\pao v1 * F) = \hardness(F) - 1$ holds.
  \end{enumerate}
  If $F \in \Usat$ and $\hardness(F) > 0$ (i.e., $\bot \notin F$), then there is a variable $v \in \var(F)$ and $\ve \in \set{0,1}$ with $\hardness(\pao v{\ve} * F) < \hardness(F)$.
\end{lem}
\begin{prf}
The assertion on the existence of $v$ and $\ve$ follows by definition. Assume now that neither of the two cases holds, i.e., that there is some $\ve \in \set{0,1}$ such that $\hardness(\pao v{\ve} * F) \le \hardness(F) - 1$ and $\hardness(\pao v{\ol{\ve}} * F) \le \hardness(F) - 2$. Consider a partial assignment $\vp$ such that $\vp * F \in \Usat$ and $\hardness(\vp * F) = \hardness(F)$ (recall Definition \ref{def:hd}). Then $v \not\in \var(\vp)$ holds. Now $\hardness(\pao v{\ve} * (\vp * F)) \le \hardness(F) - 1$ and $\hardness(\pao v{\ol{\ve}} * (\vp * F)) \le \hardness(F) - 2$, so by definition of hardness for unsatisfiable clause-sets we have $\hardness(\vp * F) \le \hardness(F) -1$, a contradiction. \Qed
\end{prf}

\begin{examp}\label{exp:hdpass}
  The simplest examples for $F \in \Cls$ with $n(F) > 0$ such that for all $x \in \lit(F)$ holds $\hardness(\pao x1 * F) = \hardness(F)$ are the elements of $\Urefc_0$. An example with hardness $1$ (in fact a Horn clause-set) is $F := \set{a \ra b, b \ra c, b \ra d, \neg c \vee \neg d} = \set{\set{\ol{a},b},\set{\ol{b},c},\set{\ol{b},d},\set{\ol{c},\ol{d}}}$ (we have $\hardness(F) > 0$ since for example the resolvent $\set{\ol{a},b} \res \set{\ol{b},c} = \set{\ol{a},c}$ is not subsumed by a clause in $F$).
\end{examp}

\subsection{Hardness under various operations}
\label{sec:hdfl}

\begin{lem}\label{lem:hdpassub}
  Consider $F \in \Cls$ and $V \sse \var(F)$. Let $P$ be the set of partial assignments $\psi$ with $\var(\psi) = V$. Then $\hardness(F) \le \abs{V} + \max_{\psi \in P} \hardness(\psi * F)$.
\end{lem}
\begin{prf}
Consider a partial assignment $\vp$ with $\vp * F \in \Usat$; we have to show $\hardness(\vp * F) \le \abs{V} + \max_{\psi \in P} \hardness(\psi * F)$. Build a resolution refutation of $\vp * F$ by first creating a splitting tree (possibly degenerated) on the variables of $V$; this splitting tree (a perfect binary tree) has height $\abs{V}$, and at each of its leaves we have a clause-set $\vp * (\psi * F)$ for some appropriate $\psi \in P$. Thus at each leaf we can attach a splitting tree of Horton-Strahler number of hardness at most $\max_{\psi \in P} \hardness(\psi * F)$, and from that (via the well-known correspondence of splitting trees and resolution trees; see \cite{Ku99b,Ku00g} for details) we obtain a resolution tree fulfilling the desired hardness bound. \Qed
\end{prf}

We obtain an upper bound on the hardness of the union of two clause-sets:
\begin{corol}\label{cor:hdunion}
  For $F_1, F_2 \in \Cls$ holds $\hardness(F_1 \cup F_2) \le \max(\hardness(F_1), \hardness(F_2)) + \abs{\var(F_1) \cap \var(F_2)}$.
\end{corol}
\begin{prf}
  Apply Lemma \ref{lem:hdpassub} with $F := F_1 \cup F_2$ and $V := \var(F_1) \cap \var(F_2)$, and apply the general upper bound $\hardness(F_1 \cup F_2) \le \max(\hardness(F_1), \hardness(F_2))$ for variable-disjoint $F_1, F_2$ (Lemma 15 in \cite{GwynneKullmann2012SlurSOFSEM}). \Qed
\end{prf}

Substitution of literals can not increase (w-)hardness:
\begin{lem}\label{lem:eqhardness}
  Consider a clause-set $F \in \Usat$ and (arbitrary) literals $x,y$. Denote by $F_{x \la y} \in \Usat$ the result of replacing $x$ by $y$ and $\ol{x}$ by $\ol{y}$ in $F$, followed by removing clauses containing complementary literals. Then we have $\hardness(F_{x \la y}) \le \hardness(F)$ and $\whardness(F_{x \la y}) \le \whardness(F)$.
\end{lem}
\begin{prf}
Consider $T: F \vdash \bot$. It is a well-known fact (and a simply exercise), that the substitution of $y$ into $x$ can be performed in $T$, obtaining $T_{x \la y}: F_{x \la y} \vdash \bot$. This is easiest to see by performing first the substitution with $T$ itself, obtaining a tree $T'$ which as a binary tree is identical to $T$, using ``pseudo-clauses'' with (possibly) complementary literals; the resolution rule for sets $C, D$ of literals with $x \in C$ and $\ol{x} \in D$ allows to derive the clause $(C \sm \set{x}) \cup (D \sm \set{\ol{x}})$, where the resolution-variables are taken over from $T$. Now ``tautological'' clauses (containing complementary literals) can be cut off from $T'$: from the root (labelled with $\bot$) go to a first resolution step where the resolvent is non-tautological, while one of the parent clauses is tautological (note that not both parent clauses can be tautological) --- the subtree with the tautological clause can now be cut off, obtaining a new pseudo-resolution tree where clauses only got (possibly) shorter (see Lemma 6.1, part 1, in \cite{Ku00g}). Repeating this process we obtain $T_{x \la y}$ as required. Obviously $\hts(T_{x \la y}) \le \hts(T)$, and if in $T$ for every resolution step at least one of the parent clauses has length at most $k$ for some fixed (otherwise arbitrary) $k \in \NNZ$, then this also holds for $T_{x \la y}$. \Qed
\end{prf}

\begin{examp}\label{exp:substhd}
  The simplest example showing for satisfiable clause-sets $F$ hardness can be increased by substitution is given by $F := \set{\set{x},\set{\ol{y}}}$ (recall $\var(x) \not= \var(y)$). Here $\hardness(F) = 0$, while $F_{x \la y} = \set{\set{y},\set{\ol{y}}}$, and thus $\hardness(F_{x \la y}) = 1$.
\end{examp}

\subsection{Bounds related to the $\rk_k$-characterisation}
\label{sec:boundsrk}

In \cite[Subsection 3.4.1]{Ku99b} a basic method to determine upper bounds on hardness for unsatisfiable clause-sets is given, where the gist is as follows:
\begin{lem}\label{lem:upperboundhd}
  Consider a class $\mc{C} \sse \Usat$ and a map $u: \mc{C} \ra \NNZ$ (like ``upper bound''). Now
  \begin{displaymath}
    \fa\,F \in \mc{C} : \hardness(F) \le u(F)
  \end{displaymath}
  holds if the following (sufficient) condition holds:
  \begin{enumerate}
  \item $\fa\, F \in \mc{C} : u(F) = 0 \Ra \hardness(F) = 0$.
  \item For $k \in \NN$, and $F \in \mc{C}$ with $u(F) =: k$ and $\hardness(F) \ge 2$ there are $x \in \lit(F)$ and $F_0, F_1 \in \mc{C}$ such that:
    \begin{enumerate}
    \item $\max(\hardness(F_0),k-1) \ge \hardness(\pao x0 * F)$ and $u(F_0) \le k-1$,
    \item $\max(\hardness(F_1),k) \ge \hardness(\pao x1 * F)$, $n(F_1) < n(F)$ and $u(F_1) \le k$.
    \end{enumerate}
  \end{enumerate}
\end{lem}
\begin{prf}
Assume for the sake of contradiction that there is $F \in \mc{C}$ with $\hardness(F) > u(F) =: k$, and consider such an $F$ with first minimal $k$ and second minimal $n(F)$. By the first condition we get $k \ge 1$, and thus $\hardness(F) \ge 2$. By minimality we have $\hardness(F_0) \le u(F_0)$ and $\hardness(F_1) \le u(F_1)$. Thus $\hardness(\pao x0 * F) \le k-1$ and $\hardness(\pao x1 * F) \le k$, whence $\hardness(F) \le k$ (Lemma \ref{lem:hdpass}). \Qed
\end{prf}

Remarks:
\begin{enumerate}
\item Assume that $\mc{C}$ is stable under application of partial assignments, and that for all $F \in \mc{C}$ and $\vp \in \Pass$ holds $u(\vp * F) \le u(F)$.
  \begin{enumerate}
  \item For case 2(a) we can also use $F_0' := \rk_{k-1}(F_0)$.
  \item For case 2(b) we can also use $F_1' := \rk_k(F_1)$.
  \end{enumerate}
\end{enumerate}

 For showing lower bounds on the hardness for unsatisfiable clause-sets, we can use the methodology developed in Subsection 3.4.2 of \cite{Ku99b}. A simplified version of Lemma 3.17 from \cite{Ku99b}, sufficient for our purposes, is as follows (with a technical correction, as explained in Example \ref{exp:lbhd}):

\begin{lem}\label{lem:lbhd}
  Consider $\mc{C} \sse \Usat$ and a function $h: \mc{C} \ra \NNZ$. Then
  \begin{displaymath}
    \fa\, F \in \mc{C} : \hardness(F) \ge h(F)
  \end{displaymath}
  if the following (sufficient) condition holds:
  \begin{enumerate}
  \item $\fa\, F \in \mc{C} : \hardness(F) = 0 \Ra h(F) = 0$.
  \item For all $F \in \mc{C}$ with $k := h(F) \ge 1$ and $v \in \var(F)$
    \begin{enumerate}
    \item either there is $\ve \in \set{0,1}$ and $F_{\ve} \in \mc{C}$ with $\hardness(F_{\ve}) \le \hardness(\pao v{\ve} * F)$, $h(F_{\ve}) \ge k$ and $n(F_{\ve}) < n(F)$,
    \item or there are $F_0, F_1 \in \mc{C}$ with $\hardness(F_{\ve}) \le \hardness(\pao v{\ve} * F)$ and $h(F_{\ve}) \ge k-1$ for both $\ve \in \set{0,1}$,
    \end{enumerate}
    or both.
  \end{enumerate}
\end{lem}
\begin{prf}
Assume for the sake of contradiction that there is $F \in \mc{C}$ with $\hardness(F) < h(F) =: k$, and consider such an $F$ with first minimal $\hardness(F)$ and second minimal $n(F)$. By the first condition we get $\hardness(F) \ge 1$.

If the first case holds, there is $\ve \in \set{0,1}$ and $F_{\ve} \in \mc{C}$ with $\hardness(F_{\ve}) \le \hardness(\pao v{\ve} * F)$ ($\le \hardness(F)$), $h(F_{\ve}) \ge k$ and $n(F_{\ve}) < n(F)$. Due to $\hardness(F_{\ve}) \le \hardness(F)$, by the minimality condition for $F$ we have $h(F_{\ve}) \le \hardness(F_{\ve})$, and thus actually $h(F_{\ve}) \le \hardness(F) < k$, contradicting the condition. So assume the second case holds.

By Lemma \ref{lem:hdpass} there is $v \in \var(F)$ and $\ve \in \set{0,1}$ with $\hardness(\pao v{\ve} * F) < \hardness(F)$. By the case assumption there is $F_{\ve} \in \mc{D}$ with $\hardness(F_{\ve}) \le \hardness(\pao v{\ve} * F)$ and $h(F_{\ve}) \le k-1$. We get $h(F_{\ve}) \le \hardness(F_{\ve})$ due to the minimality condition for $F$, while $\hardness(F_{\ve}) \le \hardness(\pao v{\ve} * F) < \hardness(F) < k$, and thus $h(F_{\ve}) \le k-2$, contradiction the condition. \Qed
\end{prf}

Lemma 3.17 in \cite{Ku99b} doesn't state for Case 2(a) the condition ``$n(F_{\ve}) < n(F)$'' from Lemma \ref{lem:lbhd}. The following example shows that this condition actually needs to be stated; fortunately in all applications in \cite{Ku99b} this (natural) condition is fulfilled.
\begin{examp}\label{exp:lbhd}
  Consider $\mc{C} := \Urefc_1 \cap \Usat$. Define $h: \mc{C} \ra \set{0,1,2}$ as $h(F) = \hardness(F)$ iff $\bot \in F$ or there is a variable $v$ with $\set{v}, \set{\ol{v}} \in F$, while otherwise $h(F) = 2$. Thus $h$ is not a lower bound on $\hardness$. We have $h(F) = 2$ if and only if $n(F) > 0$ and for all variables $v \in \var(F)$ holds $\hardness(\pao v0 * F) = 1$ or $\hardness(\pao v1 * F) = 1$. We define $F_{\ve} := \pao v{\ve} * F$ if $\hardness(\pao v{\ve} * F) < \hardness(F)$, and otherwise $F_{\ve} := F$. For $F \in \mc{C}$ with $h(F) \le 1$ trivially always Condition 2(b) is fulfilled, so consider $F \in \mc{C}$ with $h(F) = 2$, and consider $v \in \var(F)$. We do not have Condition 2(b) iff there is $\alpha \in \set{0,1}$ with $\bot \in \pao v{\alpha} * F$, but then via $\ve := \ol{\alpha}$ Condition 2(a) holds, when not considering the requirement, that the number of variables must strictly decrease.
\end{examp}

\subsection{Game characterisations}
\label{sec:hdgames}

The game of \Pudlak and Impagliazzo \cite{PI2000} is a well-known and classic Prover-Delayer game, which serves as one of the main and conceptually very simple methods to obtain resolution lower bounds for unsatisfiable formulas in CNF. The game proceeds between a Prover and a Delayer. The Delayer claims to know a satisfying assignment for an unsatisfiable clause-set, while the Prover wants to expose his lie and in each round asks for variable value. The Delayer can either choose to answer this question by setting the variable to 0/1, or can defer the choice to the Prover. In the latter case, Delayer scores one point.

This game provides a method for showing lower bounds for tree resolution. Namely, \Pudlak and Impagliazzo \cite{PI2000} show that exhibiting a Delayer strategy for a CNF $F$ that scores at least $p$ points against every Prover implies a lower bound of $2^p$ for the proof size of $F$ in tree resolution. More precisely, by Lemma \ref{lem:spacetres} we know that for unsatisfiable clause-set $F$ holds $\treespace(F) = \hardness(F) + 1$, while in \cite{EstebanToran2003CombCharacTreeRes} it was shown that the optimal value of the \Pudlak-Impagliazzo game plus one equals $\treespace(F)$, and thus $\hardness(F)$ is the optimal value of the \Pudlak-Impagliazzo game for $F$.
We remark that exactly for this reason, the game of \Pudlak and Impagliazzo does not characterise tree resolution size (precisely). In \cite{BGL13,BeyersdorffGalesiLauria2010Treelike} a modified \emph{asymmetric} version of the game is introduced, which precisely characterises tree resolution size \cite{BGL13a}.

The original \Pudlak-Impagliazzo game only works for unsatisfiable formulas.
We now show that with a variation of the game we can characterise $\hardness(F)$ for arbitrary $F$, both unsatisfiable and satisfiable. A feature of this game, not shared by the original game, is that there is just one ``atomic action'', namely the choice of a variable \emph{and} a value, for both players, and the rules are just about how this choice can be employed. This allows this game to be extended to handle also $\whardness$ (Theorem \ref{thm:whdspiel}). Delayer in both cases just extends the current partial assignment.

\begin{thm}\label{thm:hd=PI}
  Consider $F \in \Cls$. The following game is played between Prover and Delayer, where the partial assignments $\theta$ all fulfil $\var(\theta) \sse \var(F)$:
  \begin{enumerate}
  \item The two players play in turns, and Delayer starts. Initially $\theta := \epa$.
  \item A move of Delayer extends $\theta$ to $\theta' \supseteq \theta$.
  \item A move of Prover extends $\theta$ to $\theta' \supset \theta$ with $\theta' * F = \top$ or $n(\theta') = n(\theta) + 1$.
  \item The game ends as soon $\bot \in \theta * F$ or $\theta * F = \top$.
     In the first case Delayer gets as many points as variables have been assigned by Prover.
     In the second case Delayer gets zero points.
    \end{enumerate}
  Now there is a strategy of Delayer which can always achieve $\hardness(F)$ many points, while Prover can always avoid that Delayer gets $\hardness(F)+1$ or more points.
\end{thm}
\begin{prf}
  The strategy of Prover is: If $\theta * F$ is satisfiable, then extend $\theta$ to a satisfying assignment. Otherwise choose $v \in \var(F)$ and $\ve \in \set{0,1}$ such that $\hardness(\pao v{\ve} * F)$ is minimal. The strategy of Delayer is: Initially extend $\epa$ to some $\theta$ such that $\theta * F \in \Usat$ and $\hardness(\theta * F)$ is maximal. For all other moves, and also for the first move as an additional extension, as long as there are variables $v \in \var(\theta * F)$ and $\ve \in \set{0,1}$ with $\hardness(\pao v{\ve} * (\theta * F)) \le \hardness(\theta * F) - 2$, choose such a pair $(v,\ve)$ and extend $\theta$ to $\theta \cup \pao v{\ol{\ve}}$. The assertion now follows with Lemma \ref{lem:hdpass} (which is only needed for unsatisfiable $F$). \Qed
\end{prf}

Remarks:
\begin{enumerate}
\item A feature of the game of Theorem \ref{thm:hd=PI}, not shared by the game in \cite{PI2000}, is that there is just one ``atomic action'', namely the choice of a variable \emph{and} a value, for both players, and the rules are just about how this choice can be employed.
\end{enumerate}

\subsection{Characterisation by sets of partial assignments}
\label{sec:hdchpass}

We now provide an alternative characterisation of hardness of clause-sets $F$ by sets $\PP$ of partial assignments. The ``harder'' $F$ is, the better $\PP$ ``approximates'' satisfying $F$. The minimum condition is:
\begin{defi}\label{def:minconsis}
  For $F \in \Cls$ a set $\PP \sse \Pass$ is \textbf{minimal consistent} if $\var(\PP) = \bc_{\vp \in \PP} \var(\vp) \sse \var(F)$, for all $\vp \in \Pass$ holds $\bot \notin \vp * F$, and $\PP \ne \es$.
\end{defi}
$\PP$ is a \href{http://en.wikipedia.org/wiki/Glossary_of_order_theory}{partially ordered set} (by inclusion). Recall that a \emph{chain} $K$ is a subset constituting a linear order, while the \emph{length} of $K$ is $\abs{K} - 1 \in \ZZ_{\ge -1}$, and a \textbf{maximal chain} is a chain which can not be extended without breaking linearity.

\begin{defi} \label{def:weak-consistency}
  For $k \in \NNZ$ and $F \in \Usat$ let a \textbf{weakly $k$-consistent set of partial assignments for $F$} be a $\PP$ minimally consistent for $F$, such that the minimum length of a maximal chain in $\PP$ is at least $k$, and for every non-maximal $\vp \in \PP$, every $v \in \var(F) \sm \var(\vp)$ and every $\ve \in \set{0,1}$ there is $\vp' \in \PP$ with $\vp \cup \pao v\ve \sse \vp'$.
\end{defi}

Note that there might be ``gaps'' between $\vp \subset \vp'$ for $\vp, \vp' \in \PP$; this corresponds to the moves of Delayer in Theorem \ref{thm:hd=PI}, who needs to prevent all ``bad'' assignments at once.

\begin{lem}\label{lem:charPIcons}
  For all $F \in \Usat$ we have  $\hardness(F) > k$ if and only if there is a weakly $k$-consistent set for partial assignments for $F$.
\end{lem}
\begin{prf}
If there is a weakly $k$-consistent $\PP$, then Delayer from Theorem \ref{thm:hd=PI} has a strategy achieving at last $k+1$ points by choosing a minimal $\theta' \in \PP$ extending $\theta$, and maintaining in this way $\theta \in \PP$ as long as possible. And a weakly $(\hardness(F)-1)$-consistent $\PP$ for $\hardness(F) > 0$ is given by the set of all $\vp^* \in \Pass$, which are obtained from $\vp \in \Pass$ with $\bot \notin \vp * F$ by extending $\vp$ to $\vp' := \vp \cup \pao v\ve$ for such $v \in \var(F) \sm \var(\vp)$ and $\ve \in \set{0,1}$ with $\hardness(\vp' * F) = \hardness(\vp * F)$ as long as possible. \Qed
\end{prf}

\begin{quest}\label{que:weakconsun}
  For $F \in \Usat$ and $k \in \NNZ$, is the set of weakly $k$-consistent sets of partial assignments for $F$ closed under non-empty union?
\end{quest}

\subsection{Characterising depth}
\label{sec:characdep}

A similar characterisation can also be given for the depth-measure $\dep(F)$ (cf.~Definition~\ref{def:dep}). For this we relax the concept of weak consistency.

\begin{defi}
  For $k \in \NNZ$ and $F \in \Usat$ let a \textbf{very weakly $k$-consistent set of partial assignments for $F$} be a minimally consistent $\PP$ for $F$ such that $\epa \in \PP$, and for every $\vp \in \PP$ with $n(\vp) < k$ and all $v \in \var(F) \sm \var(\vp)$ there is $\ve \in \set{0,1}$ with $\vp \cup \pao v\ve \in \PP$.
\end{defi}

By \cite[Theorem 2.4]{Urquhart2011Depth} we get the following characterisation (we provide a proof due to technical differences):
\begin{lem}\label{lem:chardep}
  For all $F \in \Usat$ we have  $\dep(F) > k$ if and only if there is a very weakly $k$-consistent set for partial assignments for $F$.
\end{lem}
\begin{prf}
If $F$ has a resolution proof $T$ of height $k$, then for a weakly $k'$-consistent $\PP$ for $F$ we have $k' < k$, since otherwise starting at the root of $T$ we follow a path given by extending $\epa$ according to the extension-condition of $\PP$, and we arrive at a $\vp \in \PP$ falsifying an axiom of $T$, contradicting the definition of $\PP$. On the other hand, if $\dep(F) > k$, then there is a very weakly $k$-consistent $\PP$ for $F$ as follows: for $j \in \tb 0k$ put those partial assignments $\vp \in \Pass$ with $\var(\vp) \sse \var(F)$ and $n(\vp) = j$ into $\PP$ which do not falsify any clause derivable by a resolution tree of depth at most $k-j$ from $F$. Now consider $\vp \in \PP$ with $j := n(\vp) < k$, together with $v \in \var(F) \sm \var(\vp)$. Assume that for both $\ve \in \set{0,1}$ we have $\vp \cup \pao {v}{\ve} \notin \PP$. So there are clauses $C, D$ derivable from $T$ by a resolution tree of depth at most $k-j-1$, with $v \in C$, $\ol{v} \in D$, and $\vp * \set{C,D} = \set{\bot}$. But then $\vp * \set{C \res D} = \set{\bot}$, contradicting the defining condition for $\vp$. \Qed
\end{prf}

\section{Width-hardness}
\label{sec:hdproofcw}

We now turn to characterisations of the width-hardness measures.

\subsection{On the complexity of $k$-resolution}
\label{sec:deckres}

Since the introduction of $k$-resolution in \cite{Kl93}, the question is open whether for every (fixed) $k \in \NNZ$ and input $F \in \Cls$ the property ``$F \vdash^k \bot$''t is decidable in polynomial time. Trivially $F \vdash^0 \bot$ iff $\bot \in F$. Also for $k=1$ there is a linear-time algorithm, since $F \vdash^1 \bot$ iff $\rk_1(F) = \set{\bot}$, and for $k=2$ there is a quartic-time algorithm by \cite{BuroKleineBuening1996ResolutionShortClauses}, while nothing is known for $k \ge 3$. We can not solve this question here, but we obtain some insights into the structure of $k$-resolution refutations, which leads us to what seems the key question here.

First we need to review some known facts on \emph{input resolution}, that is, resolution trees $T: F \vdash C$ with $\hts(T) \le 1$ (that is, every node, which is not itself a leaf, has a leaf as a child), for which we write $T: F \vdash_1 C$; the axioms of $T$ are also called \emph{input clauses}. And we write $F \vdash_1 C$ if there is $C' \sse C$ and $T: F \vdash_1 C'$. Whether a clause (or a sub-clause thereof) can be derived by input resolution, is decidable in linear time:
\begin{lem}\label{lem:inpresdec}
  Consider inputs $F \in \Cls$ and $C \in \Cl$. Then $F \vdash_1 C$ iff $\rk_1(\vp_C * F) = \set{\bot}$ (where the latter is decidable in linear time).
\end{lem}
In an input resolution tree $T: F \vdash_1 C$ we call the leaves (clauses) with maximal depth (maximal distance from the root) the \emph{top clauses} of $T$ (if $T$ is trivial, then there is one top clause, otherwise we have exactly two of them).

\begin{lem}\label{lem:charackres}
  Consider $F \in \Cls$ and $k \in \NNZ$. Let $F^* := \set{C \in \Cl : F \vdash^k C \und \abs{C} \le k}$ be the set of clauses of length at most $k$ derivable from $F$ by $k$-resolution. Then $F^*$ is up to subsumption the same as the closure $F'$ of $F$ under input resolution, where at most one input clause has length $> k$, in which case it is a top clause and element of $F$. More precisely:
  \begin{enumerate}
  \item[(i)] Start with $F' := \set{C \in F : \abs{C} \le k}$.
  \item[(ii)] Assume there is a clause $C \in \Cl \sm F'$ with $\var(C) \sse \var(F)$, $\abs{C} \le k$ and a clause $D \in F$ such that there is $T: F' \cup \set{D} \vdash_1 C'$ for some $C' \sse C$, where if $D$ is an input clause of $T$, then $D$ occurs exactly once in $T$, and that as a top clause.

    Then $F' := F' \cup \set{C}$.
  \item[(iii)] Repeat this extension as long as possible, obtaining the final closure $F'$.
  \end{enumerate}
  Now we have:
  \begin{enumerate}
  \item\label{lem:charackres1} $F' \supseteq F^*$.
  \item\label{lem:charackres2} For every $C \in F'$ there is a $D \in F^*$ with $D \sse C$.
  \end{enumerate}
\end{lem}
\begin{prf}
We first prove Part \ref{lem:charackres2}, via induction on the construction of $F'$. For the beginning, as in Step (i) above, the assertion is trivial. Now consider a clause $C$ added in the process of computing $F'$ as in Step (ii) above. By induction we can replace all input clauses $E \in \premr(T)$ with clauses $E' \in F^*$ and $E' \sse E$, and obtain $T': F^* \vdash_1 C''$, $C'' \sse C'$. Now at most one input clause of $T'$ has length $> k$, and if it exists, such an input clause is a top clause of $T'$, whence for every resolution step of $T'$ at least one parent clause is of length $\le k$. It follows $C'' \in F^*$, where $C'' \sse C$, concluding the proof of Part \ref{lem:charackres2}.

For Part \ref{lem:charackres1} assume there is $C \in F^* \sm F'$. W.l.o.g.\ consider some $T: F \vdash^k C$, where for all $D \in \allcr(T) \sm \set{C}$ with $\abs{D} \le k$ we have $D \in F'$. By definition of $F'$ we have that $T$ is not trivial. Exactly one of the parent clauses of $C$ in $T$ must have length $>k$ (otherwise $C \in F'$). For this parent clause $C'$ holds $C' \notin F$, since otherwise $C \in F'$ (using $C'$ as a top clause). So consider the parent clauses of $C'$. For exactly one of them, $C''$, we have again $\abs{C''} > k$ (otherwise $C \in F'$), and furthermore we have again $C'' \notin F$ (otherwise $C \in F'$, using $C''$ as a top clause). So this process can be repeated with $C''$, which leads finally to a contradiction, since $T$ is finite. \Qed
\end{prf}

The crucial decision problem is IRES-TOP:
\begin{itemize}
\item Input $(F,C)$ with $F \in \Cls$ and $C \in \Cl$.
\item Answer YES iff there is $T: F \vdash_1 \bot$, where $C$ is a top clause of $T$, which is not used as another input clause of $T$.
\item This is equivalent to the existence of a regular $T: F \vdash_1 \bot$, where $C$ is a top clause of $T$.
\end{itemize}
In \cite{HertelUrquhart2009InputResolution}, Lemma 4.8, it is shown that for $F \in \Musat$ with $F \vdash_1 \bot$ every clause of $F$ can be used as a top clause --- however this does not apply when the top clause must not be reused (or, equivalently, if the input tree must also be regular), as the example $F := \set{\set{a},\set{\ol{a},b},\set{\ol{a},\ol{b}}}$ and $C := \set{a}$ shows: IRES-TOP for input $(F,C)$ is NO.

\begin{lem}\label{lem:suffpolytime}
  If IRES-TOP is solvable in polynomial time, then for every $k \in \NNZ$ it is decidable in polynomial time whether for input $F \in \Cls$ we have $F \vdash^k \bot$.
\end{lem}
\begin{prf}
By Lemma \ref{lem:charackres} we have $F \vdash^k \bot$ iff $\bot \in F'$. What has to be achieved is Step (ii) of the closure procedure, and this can be implemented as follows:
\begin{enumerate}
\item The outer loops runs through $C \in \Cl \sm F'$ with $\var(C) \sse \var(F)$ and $\abs{C} \le k$.
\item If $F' \vdash_1 C$ (Lemma \ref{lem:inpresdec}), then $C$ is added to $F'$.
\item Otherwise the inner loops runs through $D \in F$ with $\abs{D} > k$ and $\vp_C * \set{D} \ne \top$.
\item If IRES-TOP$(\vp_C * (F \cup \set{D}), \vp_C * D)$ yields YES, then $C$ is added to $F'$.
\end{enumerate}
This is repeated until $F'$ no longer changes, and then $\bot \in F'$ is checked. \Qed
\end{prf}

\subsection{The relation between symmetric and asymmetric width}
\label{sec:relsymmasymw}

In \cite{Ku99b,Ku00g} a stronger system than $k$-resolution was considered, which considers the closure of clause-set $F$ under input resolution, where only the conclusion is restricted to length $\le k$: If we consider Lemma \ref{lem:charackres}, then in Step (ii) all clauses of $F$ can now be used, without restriction on their position. For this stronger system polytime decision (for deriving the empty clause) follows simply with Lemma \ref{lem:inpresdec}. It is instructive to consider this system to show the basic result
\begin{displaymath}
  \wid(F) - \max(p,\whardness(F)) \le \whardness(F)
\end{displaymath}
for $F \in \Usat \cap \Pcls{p}$, which is shown in Lemma 8.5 in \cite{Ku99b}, or, more generally, in Lemma 6.22 in \cite{Ku00g}; for ease of access we give a proof here:
\begin{lem}\label{lem:inputressim}
  If for $F \in \Pcls{p}$, $p \in \NNZ$, holds $F \vdash_1 C$, then there is $T: F \vdash C'$ for some $C' \sse C$, where $\allcr(T) \in \Pcls{p'}$ for $p' := p + \abs{C}$ (i.e., all clauses in $T$ have length at most $p'$).
\end{lem}
\begin{prf}
There is an input-resolution proof of clause $C$ from clause-set $F \in \Pcls{p}$ iff $\rk_1(\vp_C * F) = \set{\bot}$, and since unit-resolution does not increase the size of clauses, we get $T_0: \vp_C * F \vdash \bot$ with $\allcr(T_0) \in \Pcls{p}$. Adding the literals of $C$ to clauses of $F$ where these literals have been eliminated by the application of $\vp_C$, we obtain $T$ as desired. \Qed
\end{prf}

\begin{corol}\label{cor:inputressim}
  For $F \in \Pcls{q}$, $q \in \NNZ$, we have $\wid(F) \le \whardness(F) + q'$, where $q' := \max(q,\whardness(F))$.
\end{corol}
\begin{prf}
W.l.o.g.\ $F \in \Usat$. We show a stronger result: Consider $k \in \NNZ$, and let $F'$ be the closure of $F \in \Pcls{q}$ under derivation via input-resolution of clauses of length at most $k$. If $\bot \in F'$, then $\wid(F) \le k + \max(q,k)$. This follows directly from Lemma \ref{lem:inputressim}, where $p := \max(q,k)$. \Qed
\end{prf}

\begin{conj}\label{con:relwidwhd}
  $\fa\, q \in \NN \, \fa\, F \in \Pcls{q} : \wid(F) \le \whardness(F) + q - 1$.
  \begin{enumerate}
  \item Holds for $q \le 2$. So the first real case is $F \in \Pcls{3} \cap \Usat$.
  \end{enumerate}
\end{conj}

\subsection{Symmetric width}
\label{sec:symw}

First it is instructive to review the characterisation for $\wid(F)$ for $F \in \Usat$ from \cite{AD2002J}, using a different formulation.
\begin{defi}\label{def:kconsissym}
  Consider $F \in \Cls$ and $k \in \NNZ$. A \textbf{symmetrically $k$-con\-sis\-tent set of partial assignments for $F$} is a minimally consistent $\PP$ for $F$, such that for all $\vp \in \PP$, all $v \in \var(F) \sm \var(\vp)$, and all $\psi \sse \vp$ with $n(\psi) < k$ there exists $\ve \in \set{0,1}$ and $\vp' \in \PP$ with $\psi \cup \pao{v}{\ve} \sse \vp'$.
\end{defi}
Remarks:
\begin{enumerate}
\item A symmetrically $k$-consistent set is also very weakly $k$-consistent.
\item Consider $F \in \Cls$ and $k \in \NNZ$. The set of symmetrically $k$-consistent set of partial assignment for $F$ is stable under non-empty union, and thus has a largest element if it is non-empty. This largest element is determined in Lemma \ref{lem:largsymkcons}.
\end{enumerate}

\begin{lem}\label{lem:kconsissym}
  Consider $F \in \Usat$ and $k \in \NNZ$. Then Duplicator wins the Boolean existential $k$-pebble game on $F$ in the sense of \cite{AD2002J} if and only if there exists a symmetrically $k$-consistent set of partial assignment for $F$.
\end{lem}
\begin{prf}
First assume that Duplicator wins the Boolean existential $k$-pebble game on $F$. Then there exists a non-empty set $\PP \sse \Pass$ with $\bot \notin \vp * F$ for all $\vp \in \PP$ fulfilling:
\begin{enumerate}
\item[(i)] for $\vp \in \PP$ we have $n(\vp) \le k$;
\item[(ii)] for $\vp \in \PP$ and $\psi \sse \vp$ we have $\psi \in \PP$;
\item[(iii)] for $\vp \in \PP$ with $n(\vp) < k$ and $v \in \var(F) \sm \var(\vp)$ there exists $\ve \in \set{0,1}$ with $\vp \cup \pao v{\ve} \in \PP$.
\end{enumerate}
Now $\PP$ is also a symmetrically $k$-consistent set of partial assignment for $F$.

In the other direction consider a symmetrically $k$-consistent set $\PP$ of partial assignment for $F$, and let $\PP'$ be the set of $\vp \in \Pass$ with $n(\vp) \le k$ such that there is $\vp' \in \PP$ with $\vp \sse \vp'$. Obviously $\PP'$ is not empty, its elements do not falsify clauses from $F$, and conditions (i), (ii) are fulfilled. It remains to consider $\vp \in \PP'$ with $n(\vp) < k$ and $v \in \var(F) \sm \var(\vp)$. There exists $\vp' \in \PP$ with $\vp \sse \vp'$. If $v \in \var(\vp')$, then by definition $\vp \cup \pao v{\vp'(v)} \in \PP'$. Otherwise there exists $\ve \in \set{0,1}$ with $\vp \cup \pao v{\ve} \in \PP'$. \Qed
\end{prf}

By Theorem 2 in \cite{AD2002J}:
\begin{corol}\label{thm:methodwhdlbsym}
  For $F \in \Usat$ and $k \in \NNZ$ holds $\wid(F) > k$ if and only if there exists a symmetrically $k$-consistent set of partial assignments for $F$.
\end{corol}

\begin{lem}\label{lem:largsymkcons}
  Consider $F \in \Usat$ and $k \in \NNZ$ with $\wid(F) > k$. Then the largest symmetrically $k$-consistent set of partial assignments for $F$ is the set of partial assignments $\vp \in \Pass$ with $\bot \notin \vp * F^*$, where $F^*$ is the closure under symmetric $k$-resolution, that is, the set of all clauses $C$ such that there a resolution tree $T: F \vdash C$, where $\allcr(T) \in \Pcls{k}$ (compare Definition \ref{def:swid}).
\end{lem}
\begin{prf}
Let $\PP_0$ be the set of partial assignments $\vp \in \Pass$ with $\bot \notin \vp * F^*$. By the proof of Theorem 2 in \cite{AD2002J} we know that $\PP_0$ is symmetrically $k$-consistent for $F$. Now assume that there is a symmetrically $k$-consistent set $\PP$ for $F$ with $\PP \not\sse \PP_0$. So there is $E' \in F^*$ with $\vp * \set{E} = \set{\bot}$ for some $\vp \in \PP$. Consider some $T : F \vdash E'$ with $\allcr(T) \in \Pcls{k}$. Since $\bot \notin \vp * F$ for all $\vp \in \PP$, there is a resolution step $E = C \res D$ in $R$, such that $\bot \notin \vp * \set{C,D}$ for all $\vp \in \PP$, but $\vp * \set{E} = \set{\bot}$ for some $\vp \in \PP$. Consider the resolution literal $C \cap \ol{D} = \set{x}$. We have $\var(x) \notin \var(\vp)$, since otherwise $\vp * \set{E} = \set{\bot}$. Let $\psi$ be the restriction of $\vp$ to $\var(C) \sm \set{\var(x)}$. By definition and w.l.o.g.\ there is $\vp' \in \PP$ with $\psi \cup \pao x0 \sse \vp'$. But then $\vp * \set{C} = \set{\bot}$. \Qed
\end{prf}

\subsection{Characterisation by sets of partial assignments}
\label{sec:whdcharpass}

Similar to Definition \ref{def:kconsissym}, we characterise asymmetric width --- the only difference is, that the extensions must work for both truth values.
\begin{defi}\label{def:kconsis}
  Consider $F \in \Cls$ and $k \in \NNZ$. A \textbf{$k$-consistent set of partial assignments for $F$} is a minimally consistent $\PP$ for $F$, such that for all $\vp \in \PP$, all $v \in \var(F) \sm \var(\vp)$, all $\psi \sse \vp$ with $n(\psi) < k$ and for both $\ve \in \set{0,1}$ there is $\vp' \in \PP$ with $\psi \cup \pao{v}{\ve} \sse \vp'$.
\end{defi}
Some remarks:
\begin{enumerate}
\item An equivalent formulation is:
  \begin{enumerate}
  \item $\PP \not= \es$.
  \item $\PP$ is stable under subset-formation, that is, for $\vp \in \PP$ and $\psi \in \Pass$ with $\psi \subset \vp$ holds $\psi \in \PP$.
  \item For all $\vp \in \PP$ holds $\bot \notin \vp * F$.
  \item For all $\vp \in \PP$ and $v \in \var(F) \sm \var(\vp)$ holds:
    \begin{enumerate}
    \item There is $\psi \in \PP$ with $\vp \subset \psi$ and $\var(\psi) = \var(\vp) \cup \set{v}$, or
    \item For all $\psi \sse \vp$ with $n(\psi) < k$ and for both $\ve \in \set{0,1}$ holds $\psi \cup \pao{v}{\ve} \in \PP$.
    \end{enumerate}
  \end{enumerate}
  Given $\PP$ as in Definition \ref{def:kconsis}, these conditions are fulfilled by adding all sub-partial-assignments, while in the other direction all maximal elements are selected.
\item If $\PP$ is $k$-consistent for $F$ and $F' \sse F$, then $\PP$ is also $k'$-consistent for $F'$ for all $0 \le k' \le k$.
\item $0$-consistency:
  \begin{enumerate}
  \item Any $\PP \sse \Pass$ is $0$-consistent for any $F \in \Cls$ iff $\PP \ne \es$ and for all $\vp \in \PP$ holds $\bot \notin \vp * F$.
  \item $\set{\epa}$ is $0$-consistent for any $F \in \Cls$ iff $\bot \notin F$.
  \end{enumerate}
\item $\Pass$ is $k$-consistent for any $F \in \Cls$ and any $k \in \NNZ$ iff $F = \top$.
\item Every $k$-consistent set of partial assignments for $F$ is also symmetrically $k$-consistent for $F$ (Definition \ref{def:kconsissym}).
\item Consider $F \in \Cls$ and $k \in \NNZ$. The set of $k$-consistent set of partial assignment for $F$ is stable under non-empty union, and thus has a largest element if it is non-empty. This largest element is determined in Lemma \ref{lem:largkcons}.
\end{enumerate}

\begin{examp}\label{exp:kconsis}
  Consider $F \in \Cls \sm \set{\top}$ and $k \in \NNZ$ such that for all $C \in F$ holds $\abs{C} > k$. Then the set of all $\vp \in \Pass$ with $\var(\vp) \sse \var(F)$ and $n(\vp) = k$ is $k$-consistent.
\end{examp}

Similarly to \cite[Theorem 2]{AD2002J}, where the authors provide a characterisation of $\wid(F)$, we obtain a characterisation of asymmetric width-hardness:
\begin{thm}\label{thm:methodwhdlb}
  For $F \in \Usat$ and $k \in \NNZ$ holds $\whardness(F) > k$ if and only if there exists a $k$-consistent set of partial assignments for $F$.
\end{thm}
\begin{prf}
First assume $\whardness(F) > k$. Let $F' := \set{C \in \Cl \mb \ex\, R : F \vdash^k C}$. Note that by definition $F \sse F'$, while by assumption we have $\bot \notin F'$. Now let
\begin{displaymath}
  \PP := \set{\vp \in \Pass : \bot \notin \vp * F'}.
\end{displaymath}
Note that for $\vp \in \PP$ and $\psi \sse \vp$ we have $\psi \in \PP$. We show that $\PP$ is a $k$-consistent set of partial assignments for $F$. Consider $\vp \in \PP$, $v \in \var(F) \sm \var(\vp)$ and $\psi \sse \vp$ with $n(\psi) < k$. Assume that there is $\ve \in \set{0,1}$, such that for $\psi' := \psi \cup \pao{v}{\ve}$ there is no $\vp' \in \PP$ with $\psi' \sse \vp'$. Thus there is $E \in F'$ with $\psi' * \set{E} = \set{\bot}$; so we have $v \in \var(E)$ and $\abs{E} \le k$. Now $E$ is resolvable with either $C$ or $D$ via $k$-resolution, and for the resolvent $R \in F'$ we have $\vp * \set{R} = \set{\bot}$ contradicting the definition of $\PP$.

Assume that $\PP$ is a $k$-consistent set of partial assignments for $F$. For the sake of contradiction assume there is $T : F \vdash^k \bot$. We show by induction on $\height_T(w)$ that for all $w \in \nds(T)$ and all $\vp \in \PP$ holds $\vp * \set{C(w)} \not= \set{\bot}$, which at the root of $T$ (where the clause-label is $\bot$) yields a contradiction. If $\height_T(w) = 0$ (i.e., $w$ is a leaf), then the assertion follows by definition; so assume $\height_T(w) > 0$. Let $w_1, w_2$ be the two children of $w$, and let $C := C(w)$ and $C_i := C(w_i)$ for $i \in \set{1,2}$. W.l.o.g.\ $\abs{C_1} \le k$. Note $C = C_1 \res C_2$; let $v$ be the resolution variable, where w.l.o.g.\ $v \in C_1$. Consider $\vp \in \PP$; we have to show $\vp * \set{C} \not= \set{\bot}$, and so assume $\vp * \set{C} = \set{\bot}$. By induction hypothesis we know $\bot \notin \vp * \set{C_1, C_2}$, and thus $v \notin \var(\vp)$. Let $\psi := \vp \rstr (\var(C_1) \sm \set{v})$, and $\psi' := \psi \cup \pao v0$. There is $\vp' \in \PP$ with $\psi' \sse \vp'$, thus $\psi' * \set{C_1} = \set{\bot}$ contradicting the induction hypothesis. \Qed
\end{prf}

\begin{lem}\label{lem:largkcons}
  Consider $F \in \Usat$ and $k \in \NNZ$ with $\whardness(F) > k$. Then the largest $k$-consistent set of partial assignments for $F$ is the set of partial assignments $\vp \in \Pass$ with $\bot \notin \vp * F^*$, where $F^*$ is the closure under $k$-resolution, that is, the set of all clauses $C$ such that there a resolution proof $R : F \vdash^k C$.
\end{lem}
\begin{prf}
Let $\PP_0$ be the set of partial assignments $\vp \in \Pass$ with $\bot \notin \vp * F^*$. By the proof of Theorem \ref{thm:methodwhdlb} we know that $\PP_0$ is $k$-consistent for $F$. Now assume that there is a $k$-consistent set $\PP$ for $F$ with $\PP \not\sse \PP_0$. So there is $E' \in F^*$ with $\vp * \set{E} = \set{\bot}$ for some $\vp \in \PP$. Consider some $R : F \vdash^k E'$. Since $\bot \notin \vp * F$ for all $\vp \in \PP$, there is a resolution step $E = C \res D$ in $R$, such that $\bot \notin \vp * \set{C,D}$ for all $\vp \in \PP$, but $\vp * \set{E} = \set{\bot}$ for some $\vp \in \PP$. Consider the resolution literal $C \cap \ol{D} = \set{x}$. We have $\var(x) \notin \var(\vp)$, since otherwise $\vp * \set{E} = \set{\bot}$. W.l.o.g.\ $\abs{C} \le k$. Let $\psi$ be the restriction of $\vp$ to $\var(C) \sm \set{\var(x)}$. By definition there is $\vp' \in \PP$ with $\psi \cup \pao x0 \sse \vp'$. But then $\vp * \set{C} = \set{\bot}$. \Qed
\end{prf}

\subsection{Game characterisation}
\label{sec:whdgame}

The characterisation of asymmetric width by partial assignments from the previous subsection will now be employed for a game-theoretic characterisation; in fact, the $k$-consistent set of partial assignments will directly translate into winning strategies. We only handle the unsatisfiable case here --- the general case can be handled as in Theorem \ref{thm:hd=PI}.

\begin{thm}\label{thm:whdspiel}
  Consider $F \in \Usat$. The following game is played between Prover and Delayer (as in Theorem \ref{thm:hd=PI}, always $\var(\theta) \sse \var(F)$ holds):
  \begin{enumerate}
  \item The two players play in turns, and Delayer starts. Initially $\theta := \epa$.
  \item Delayer extends $\theta$ to $\theta' \supseteq \theta$.
  \item Prover chooses some $\theta'$ compatible with $\theta$ such that $\abs{\var(\theta') \sm \var(\theta)} = 1$.
  \item If $\bot \in \theta * F$, then the game ends, and Delayer gets the maximum of $n(\theta')$ chosen by Prover as points ($0$ if Prover didn't make a choice).
  \item Prover must play in such a way that the game is finite.
  \end{enumerate}
  We have the following:
  \begin{enumerate}
  \item\label{thm:whdspiel1} For a strategy of Delayer, which achieves $k \in \NN$ points whatever Prover does, we have $\whardness(F) \ge k$.
  \item\label{thm:whdspiel2} For a strategy of Prover, which guarantees that Delayer gets at most $k \in \NNZ$ points in any case, we have $\whardness(F) \le k$.
  \item\label{thm:whdspiel3} There is a strategy of Delayer which guarantees at least $\whardness(F)$ many points (whatever Prover does).
  \item\label{thm:whdspiel4} There is a strategy of Prover which guarantees at most $\whardness(F)$ many points for Delayer (whatever Delayer does).
  \end{enumerate}
\end{thm}
\begin{prf}
W.l.o.g.\ $\bot \notin F$. Part \ref{thm:whdspiel1} follows by Part \ref{thm:whdspiel4} (if $\whardness(F) < k$, then Prover could guarantee at most $k-1$ points), and Part \ref{thm:whdspiel2} follows by Part \ref{thm:whdspiel3} (if $\whardness(F) > k$, then Delayer could guarantee at least $k+1$ points).

Let now $k := \whardness(F)$.
For Part \ref{thm:whdspiel3}, a strategy of Delayer guaranteeing $k$ many points (at least) is as follows: Delayer chooses a $(k - 1)$-consistent set $\PP$ of partial assignment (by Theorem \ref{thm:methodwhdlb}). The move of Delayer is to choose some $\theta' \in \PP$. If Prover then chooses some $\theta'$ with $n(\theta') \le k - 1$, then the possibility of extension is maintained for Delayer. In this way the empty clause is never created. Otherwise the Delayer has reached his goal, and might choose anything.

It remains to show that Prover can force the creation of the empty clause such that Delayer obtains at most $k$ many points. For that consider a resolution refutation $R: F \vdash \bot$ which is a $k$-resolution tree. The strategy of Prover is to construct partial assignments $\psi$ (from $\theta$ as given by Delayer) which falsify some clause $C$ of length at most $k$ in $R$, where the height of the node is decreasing --- this will falsify finally some clause in $F$, finishing the game. The Prover considers initially (before the first move of Delayer) just the root. When Prover is to move, he considers a path from the current clause to some leaf, such that only clauses of length at most $k$ are on that path. There must be a first clause $C$ (starting from the falsified clause, towards the leaves) on that path not falsified by $\theta$ (since $\theta$ does not falsify any axiom). It must be the case that $\theta$ falsifies all literals in $C$ besides one literal $x \in C$, where $\var(x) \notin \var(\theta)$. Now Prover chooses $\psi$ as the restriction of $\theta$ to $\var(C) \sm \set{\var(x)}$ and extends $\psi$ by $x \ra 0$. \Qed
\end{prf}

We already remarked that always $\whardness(F) \le \hardness(F)$. Based on the game characterisations shown here we provide an easy alternative proof for this fundamental fact for $F \in \Usat$.
\begin{lem}\label{lem:althdg}
  Consider the game of Theorem \ref{thm:whdspiel}, when restricted in such a way that Prover must always choose some $\theta'$ with $n(\theta') > \theta'_0$, where $\theta'_0$ is the choice of Prover in the previous round. Then this game is precisely the game of Theorem~\ref{thm:hd=PI} (characterising hardness).
\end{lem}

\begin{corol}\label{lem:whdhdg}
  For all $F \in \Cls$ we have $\whardness(F) \le \hardness(F)$.
\end{corol}

\begin{quest}\label{que:characgss}
  Is there a similar game-characterisation of $\semspace(F)$ ? And what about $\resspace(F)$ ?
\end{quest}

\begin{quest}\label{que:whdchsat}
  Can the characterisation of Theorem \ref{thm:methodwhdlb} be generalised (in a natural way) to all clause-sets?
  \begin{enumerate}
  \item It seems that an initial round is needed, choosing a hardest unsatisfiable sub-instance, and this can not be reasonably handled by such ``consistent'' sets of partial assignments.
  \end{enumerate}
\end{quest}

\begin{quest}\label{que:whdregchar}
  It should be possible to adapt Theorem \ref{thm:whdspiel} to regular resolution width, as considered in \cite{Urquhart2012RegularWidth} (there characterised by a variation on the game).
\end{quest}

\subsection{SAT solving}
\label{sec:widSAT}

\begin{quest}\label{que:widthsatbound}
  In \cite{AtseriasFichteThurley2009ClauseLearningBoundedWidth} it is shown that CDCL solvers with certain restart strategies can polynomially simulate symmetric-width restricted resolution, with asymptotic bounds on the runtime and number of restarts of the solver. Can we demonstrate similar bounds for the asymmetric-width?
\end{quest}

\section{Semantic space}
\label{sec:widsemsp}

We have already seen in Theorem~\ref{lem:whdhdg}, that our game-theoretic characterisations allow quite easy and elegant proofs on tight relations between different hardness measures. Our next result also follows this paradigm. It provides  a striking relation between asymmetric width and semantic space. We recall that Atserias and Dalmau \cite[Theorem 3]{AD2002J} have shown $\wid(F) \leq \resspace(F) + r-1$, where $F \in \Usat \cap \Pcls{r}$ (all $r \ge 0$ are allowed; note that now we can drop the unsatisfiability condition). We generalise this result in Theorem~\ref{thm:resspace} below, replacing resolution space $\resspace(F)$ by the tighter notion of semantic space $\semspace(F)$. More important, we eliminate the additional $r-1$ in the inequality, by changing symmetric width $\wid(F)$ into asymmetric width $\whardness(F)$ (cf.\ Lemma~\ref{cor:inputressim} for the relation between these two measures). First a lemma similar to \cite[Lemma 5]{AD2002J}:
\begin{lem}\label{lem:conssemsp}
  Consider $F \in \Cls$, a $k$-consistent set $\PP$ of partial assignments for $F$ ($k \in \NNZ$), and a semantic $k$-sequence $F_1,\dots,F_p$ for $F$ (recall Definition \ref{def:spaceres}). Then there exist $\vp_i \in \PP$ with $\vp_i * F_i = \top$ for each $i \in \tb 1p$.
\end{lem}
\begin{prf}
  Set $\vp_1 := \epa \in \PP$. For $i \in \tb 2p$ the partial assignment $\vp_i$ is defined inductively as follows. If $\vp_{i-1} * F_i = \top$, then $\vp_i := \vp_{i-1}$; this covers the case where $F_i$ is obtained from $F_{i-1}$ by addition of inferred clauses and/or removal of clauses. So consider $F_i = F_{i-1} \cup \set{C}$ for $C \in F \sm F_{i-1}$ (thus $c(F_i) < k$), and we assume $\vp_{i-1} * F_i \not= \top$. So there is a literal $x \in C$ with $\var(x) \notin \vp_{i-1}$, since $\vp_{i-1}$ does not falsify clauses from $F$. Choose some $\psi \sse \vp_{i-1}$ with $n(\psi) \le c(F_{i-1})$ such that $\psi * F_{i-1} = \top$.\footnote{For every partial assignment $\vp$ and every clause-set $F$ with $\vp * F = \top$ there exists $\psi \sse \vp$ with $n(\psi) \sse c(F)$ and $\psi * F = \top$; see for example Lemma 4 in \cite{AD2002J}, and see Corollary 8.6 in \cite{Kullmann2007ClausalFormZI} for a generalisation.} By the third condition from Definition \ref{def:kconsis} there is $\vp_i \in \PP$ with $\psi \cup \pao x1 \sse \vp_i$, whence $\vp_i * F_i = \top$. \Qed
\end{prf}

If the sequence $F_1,\dots,F_p$ is only a semantic $k+1$-sequence, then we can not find satisfiable assignments in $\PP$ for all $F_i$ in general:
\begin{examp}\label{exp:conssemsp}
  Let $k := 0$, $F := \set{\set{v}}$, $\PP := \set{\epa}$, and let $p := 2$. Then $(\top,F)$ is a semantic $k+1$-sequence for $F$, while $\PP$ is a $0$-consistent set of partial assignments for $F$, and there is no $\vp \in \PP$ with $\vp * F = \top$.
\end{examp}

\begin{quest}\label{que:conssemsp}
  Does it hold for complete semantic $(k+1)$-sequences for $F$, that there can not be $k$-consistent sets of partial assignments for $F$ ?
  \begin{enumerate}
  \item If in Lemma \ref{lem:conssemsp} the $k$-consistency would be really needed, then for all $F_i$ with $c(F_i) = k$ such that $F_i$ is obtained by axiom-download we had that $F_i$ is matching-satisfiable (see \cite{Kullmann2007HandbuchMU}). Does this help? Can we avoid such $F_i$ in general?
  \end{enumerate}
\end{quest}

We can now show the promised generalisation of \cite[Theorem 3]{AD2002J}:
\begin{thm}\label{thm:resspace}
  For $F \in \Usat$ holds $\whardness(F) \le \semspace(F)$.
\end{thm}
\begin{prf}
Assume $\whardness(F) > \semspace(F)$; let $k := \semspace(F)$. By Theorem \ref{thm:methodwhdlb} follows the existence of a $k$-consistent set $\PP$ of partial assignments for $F$. Let $(F_1, \dots, F_p)$ be a complete semantic $k$-sequence for $F$ according to Definition \ref{def:spaceres}. Now for the sequence $(\vp_1,\dots,\vp_p)$ according to Lemma \ref{lem:conssemsp} we have $\vp_p * F_p = \top$, contradicting $F_p \in \Usat$. \Qed
\end{prf}

\begin{corol}\label{cor:hierspace}
  For all $F \in \Cls$ holds
  \begin{displaymath}
    \whardness(F) \le \semspace(F) \le \resspace(F) \le \treespace(F) = \hardness(F)+1.
  \end{displaymath}
\end{corol}

We conclude by an application of the extended measures $\treespace, \semspace: \Cls \ra \NNZ$. In \cite{GwynneKullmann2013GoodRepresentationsI} it is shown that for every $k$ there are clause-sets in $\Urefc_{k+1}$ where every (logically) equivalent clause-set in $\Wrefc_k$ is exponentially bigger. This implies, in the language of representing boolean functions via CNFs, that allowing the tree-space to increase by $2$ over semantic space allows for an exponential saving in size (regarding logical equivalence):
\begin{corol}\label{cor:hierspace2}
  For $k \in \NN$ there is a sequence $(F_n)$ of clause-sets with $\treespace(F_n) \le k+2$, where all equivalent $(F_n')$ with $\semspace(F_n') \le k$ are exponentially bigger.
\end{corol}

\begin{quest}\label{que:whd2ss}
  Do there exist $F \in \Usat$ with $\whardness(F) = 2$ and $\semspace(F)$ arbitrary high?
  \begin{enumerate}
  \item How to show lower bounds on semantic space different from $\whardness$ ?
  \end{enumerate}
\end{quest}

\begin{conj}\label{con:improvesslb}
  We actually have $\whardness(F) + 1 \le \semspace(F)$ for $F \in \Usat$.
  \begin{enumerate}
  \item This would be the case if Question \ref{que:conssemsp} would have a positive answer.
  \item See Question \ref{que:spacephp} for an application.
  \item Can we somehow show, that in the context of semantic sequences for appropriate $F \in \Cls$ and $C \in \primec_0(F)$ there is a $(c(F)-1)$-resolution proof of $C$ ?!
  \begin{enumerate}
  \item The statement holds for $C = \bot$ and arbitrary $F$.
  \item Perhaps $F$ and $C$ can be amended in semantic sequences so that we get the statement?
  \item The goal is to show that if there is a complete semantic $k$-sequence for $F$, then we can construct from that a $(k-1)$-resolution refutation of $F$.
  \end{enumerate}
\end{enumerate}
\end{conj}

\begin{quest}\label{que:ssspass}
  Can the notion of a $k$-consistent set $\PP$ of partial assignments (Definition \ref{def:kconsis}) for $F \in \Usat$ be weakened, so that $\semspace(F)>k$ holds iff such a set of partial assignments exists?
  \begin{enumerate}
  \item The proof of Lemma \ref{lem:conssemsp} should be key. $\PP$ would have the property that for every semantic $k$-sequence $F_1,\dots,F_p$ for $F$ there is a $\vp \in \PP$ with $\vp * F_p = \top$.
  \item Instead of asking that all $\psi \sse \vp$ in Definition \ref{def:kconsis} with $n(\psi) < k$ can be extended, only ``relevant'' such $\psi$ should be considered. But that seems hard to do, since only the partial assignments are at hand?
  \item Perhaps one could ask for $\vp \in \PP$ and a clause $C \in F$ with $\vp * \set{C} \ne \top$, that there is $\vp' \in \PP$ with $\vp \subset \vp'$ and $\vp' * \set{C} = \top$ ? But that is too strong, since it doesn't depend on $k$. And a suitable sub-assignment $\psi$ of $\vp$ needs to be extended, not $\vp$ itself. That suitable sub-assignment is one satisfying a clause-set $G$ with $k-1$ clauses. One could restrict attention to $G$ with $F \models G$ --- does this help?
  \item The criterion $\vp \in \PP \leadsto \vp' \in \PP$ could thus be: For $\vp \in \PP$ and every clause-set $G$ with $\vp * G = \top$, $c(G) < k$, $\var(G) \sse \var(F)$ and $F \models G$ and every $C \in F$ with $\vp * \set{C} \ne \top$ there exists $\psi \sse \vp$ with $\psi * G = \top$ and $\vp' \in \PP$ with $\psi \subset \vp'$ and $\vp' * \set{C} = \top$.
  \item More radically, one doesn't need $\psi$: For $\vp \in \PP$ and every clause-set $G$ with $\vp * G = \top$, $c(G) < k$, $\var(G) \sse \var(F)$ and $F \models G$ and every $C \in F$ there exists $\vp' \in \PP$ with $\vp' * (G \cup \set{C}) = \top$.
  \end{enumerate}
\end{quest}

\subsection{SAT solving}
\label{sec:ssSAT}

\begin{quest}\label{que:2xor}
  What is the semantic space of the (unsatisfiable) example of 2 XOR-clauses (Theorem \ref{thm:2xor})?
\end{quest}

\section{Blocked clauses}
\label{sec:blcl}

``Blocked clauses'', a form of redundant clauses, were introduced in \cite{Ku96a,Ku97b}, as a means of adding ``valuable'' and removing ``superfluous'' clauses for SAT-solving. For a recent overview on their use in SAT-solving see \cite{JarvisaloBiereHeule2012BlockedClauseElimination} (while a general framework for showing correctness of adding and removing clauses for clause-learning SAT-solvers has been developed in \cite{JaervisaloHeuleBiere2012Inprocessing}. The proof-theoretic study of blocked clauses was started in \cite{Ku96c}.
\begin{defi}\label{def:blockKl}
  A clause $C \in \Cl$ is \textbf{blocked for $x \in C$ w.r.t.\ $F \in \Cls$}, if for all $D \in F$ with $\ol{x} \in D$ holds $\abs{C \cap \ol{D}} \ge 2$. And $C$ is \textbf{blocked w.r.t.\ $F$}, if there is $x \in C$, such that $C$ is blocked for $x$ w.r.t.\ $F$.
\end{defi}
Remarks:
\begin{enumerate}
\item A clause $C$ is blocked for $x$ w.r.t.\ a clause-set $F$ iff $C$ can not be resolved on $x$ with any clause from $F$.
\item A literal $x$ is pure for $F \in \Cls$ iff $\set{x}$ is blocked w.r.t.\ $F$.
\item Adding a clause $C$ to $F$, where $C$ is blocked w.r.t.\ $F$, results in a clause-set satisfiability-equivalent to $F$.
\end{enumerate}

\subsection{Extended Resolution}
\label{sec:ER}

\begin{defi}\label{def:exresproof}
  A \textbf{restricted extension} of a clause-set $F$ is a clause-set $F' \supseteq F$ obtained from $F$ by repeated application of
  \begin{displaymath}
    F \leadsto F \cup \setb{ \set{\ol{x},a,b}, \set{x,\ol{a}}, \set{x, \ol{b}} }
  \end{displaymath}
  where
  \begin{itemize}
  \item $x$, $a$, $b$ are literals,
  \item $\var(\set{a,b}) \sse \var(F)$, $\var(a) \not= \var(b)$,
  \item $\var(x) \notin \var(F)$.
  \end{itemize}
  A \textbf{extension} is obtained from a clause-set $F$ by repeated applications of
  \begin{displaymath}
    F \leadsto F \cup E,
  \end{displaymath}
  where $E = \primec_0(x \lra f)$ for some boolean function $f$ and literal $x$, such that
  \begin{enumerate}
  \item $\var(f) \sse \var(F)$
  \item $\var(x) \notin \var(F)$.
  \end{enumerate}
  A \textbf{(restricted) extended resolution proof} of a clause $C$ from clause-set $F$ is a resolution proof of $C$ from a (restricted) extension of $F$. Regarding complexity we use for a clause-set $F$ the following notions:
  \begin{itemize}
  \item the \textbf{extended-resolution complexity} $\bmm{\compex(F)} \in \NN$ is the minimum of $\compr(F')$ for extensions $F'$ of $F$;
  \item the \textbf{extended-tree-resolution complexity} $\bmm{\comptex(F)} \in \NN$ is the minimum of $\comptr(F')$ for extensions $F'$ of $F$.
  \end{itemize}
\end{defi}
Remarks:
\begin{enumerate}
\item A restricted extension is an extension using $x \lra (a \oder b)$.
\item See Subsection \ref{sec:blcler} for a combinatorial generalisation via ``blocked clauses''.
\end{enumerate}

\subsection{Blocked clauses and Extended Resolution}
\label{sec:blcler}

In \cite{Ku96c} blocked clauses were considered as a generalisation of extended resolution:
\begin{lem}\label{lem:erviablcl}
  Consider a one-step extension $F' = F \cup E$ of $F \in \Cls$ according to Definition \ref{def:exresproof}.
  \begin{enumerate}
  \item All clauses $C \in E$ must contain $\var(x)$, and are furthermore blocked for this literal w.r.t.\ $E$
  \item So all these clauses can be added, in any order, as blocked clauses (always w.r.t.\ the current extended clause-set).
  \end{enumerate}
\end{lem}
\begin{prf}
  XXX  from $x \lra f$ only clauses containing $\var(x)$ follow XXX
\end{prf}

XXX some overview on \cite{Ku96c} XXX

\section{Application: PHP}
\label{sec:hd}

\subsection{Fundamental definitions}
\label{sec:phpdef}

The \textbf{pigeon-hole principle} states that there is an injective map from $\tb1m$ to $\tb1k$ for $m,k \in \NNZ$ iff $m \le k$. So when putting $m$ pigeons into $k$ holes, if $m > k$ then at least one hole must contain two or more pigeons. We formalise the pigeon-hole principle as a clause-set \bmm{\php^m_k}, which is unsatisfiable iff $m > k$.
\begin{defi}\label{def:php}
  We use variables $p_{i,j} \in \Va$ for $i, j \in \NN$ such that $(i,j) \not= (i',j') \Ra p_{i,j} \not= p_{i',j'}$. For $m, k \in \NNZ$ let
  \begin{eqnarray*}
    F^{\ge 1} & := & \setb{\set{p_{i,j} \mb j \in \tb1k}}_{i \in \tb1m }\\
     F_{\ge 1} & := & \setb{\set{p_{i,j} \mb i \in \tb 1m}}_{j \in \tb 1k }\\
    F^{\le 1} & := & \setb{\set{\ol{p_{i,j_1}},\ol{p_{i,j_2}}} \mb i \in \tb 1m, j_1, j_2 \in \tb 1k, j_1 \not= j_2}\\
    F_{\le 1} & := & \setb{ \set{\ol{p_{i_1,j}},\ol{p_{i_2,j}}} \mb i_1,i_2 \in \tb1m, i_1 \not= i_2, j \in \tb1k }.
  \end{eqnarray*}
  The \textbf{pigeon-hole clause-set} $\bmm{\php^m_k} \in \Cls$ for $m,k \in \NNZ$ uses variables $p_{i,j}$ for $i \in \tb 1m$, $j \in \tb 1k$, and is defined, together with the \textbf{functional}, \textbf{onto}, and \textbf{bijective} form, as
  \begin{eqnarray*}
    \bmm{\php^m_k} & := & F^{\ge 1} \cup F_{\le 1} \in \Cls\\
    \bmm{\fphp^m_k} & := & \php^m_k \cup F^{\le 1} \in \Cls\\
    \bmm{\ophp^m_k} & := & \php^m_k \cup F_{\ge 1} \in \Cls\\
    \bmm{\ofphp^m_k} & := & \php^m_k \cup F^{\le 1} \cup F_{\ge 1} \in \Cls.
  \end{eqnarray*}
\end{defi}
Note that $\ofphp^m_k$ is isomorphic to $\ofphp^k_m$, where the isomorphism maps variables $p_{i,j}$ to $p_{j,i}$. We have $n(\php^m_k) = n(\fphp^m_k) = n(\ophp^m_k) = n(\ofphp^m_k) = m \cdot k$, and
\begin{eqnarray*}
  c(F^{\ge 1}) & = & m\\
  c(F_{\ge 1}) & = & k\\
  c(F^{\le 1}) & = & m \cdot \binom k2\\
  c(F_{\le 1}) & = & k \cdot \binom m2.
\end{eqnarray*}
Furthermore
\begin{enumerate}
\item $\php^m_k \in \Sat \Lra \fphp^m_k \in \Sat \Lra m \le k$,
\item $\ophp^m_k \in \Sat \Lra (m \le k \und m=0 \Ra k=0)$,
\item $\ofphp^m_k \in \Sat \Lra m = k$.
\item $\php^m_m, \fphp^m_m, \ophp^m_m, \ofphp^m_m$ are all equivalent to the boolean function on variables $p_{i,j}$ (with $i, j \in \tb 1m$), which is true iff the corresponding bipartite graph is a perfect matching, where that graph is obtained by interpreting $p_{i,j}$ as having an edge or not connecting vertex $i$ and vertex $j$.
\end{enumerate}

In order to determine hardness of satisfiable pigeon-hole principles, we determine their prime implicates:
\begin{quest}\label{que:prcphp}
  There should be a lemma of the content: If $\vp * \php^m_k \in \Usat$, then except of trivial cases there is a $\php^{m'}_{k'}$ for $m' \le m$, $k' \le k$ and $m' > k$ embedded into $\vp * \php^m_k$. If $\vp$ is minimal, then there should be exactly one such sub-php.

  This should help with proving Lemma \ref{lem:prcpig}.
\end{quest}

The prime implicates of (satisfiable) pigeonhole principles:
\begin{lem}\label{lem:prcpig}
  For $m, k \in \NNZ$ holds:
  \begin{enumerate}
  \item If $m \le k$ then $\primec_0(\php^m_k) = $ XXX we have $\php^m_k \sse \primec_0(\php^m_k)$ XXX plus precisely the minimal assignments which result, up to isomorphism, in some $\php^{m'}_{k'}$ for $0 \le m' \le m$ and $0 \le k' < m'$ XXX
  \item If $m \le k$ then $\primec_0(\fphp^m_k) = $ XXX we have $\fphp^m_k \sse \primec_0(\fphp^m_k)$ XXX
  \item If $0 < m \le k$ then $\primec_0(\ophp^m_k) = $ XXX we have $\ophp^m_k \sse \primec_0(\ophp^m_k)$ XXX
  \item If $m = k$ then $\primec_0(\ofphp^m_k) = $ XXX we have $\ofphp^m_k \sse \primec_0(\ofphp^m_k)$ XXX
  \end{enumerate}
\end{lem}
\begin{prf}
  XXX
\end{prf}

\subsection{Hardness}
\label{sec:phphd}

Strengthening Lemma 6 in \cite{AD2002J} (where we now don't need to consider the ``standard non-deterministic extension'' in order to get rid off the long clauses):
\begin{lem}\label{lem:uphpwhardness}
  For all $m, k \in \NNZ$ with $m > k$ we have $\whardness(\ofphp^m_k) \ge k$.
\end{lem}
\begin{prf}
For $k=0$ we have $\bot \in \ofphp^m_k$, and thus $\whardness(\ofphp^m_k) = 0$. So assume $k \ge 1$. Let $\PP \subset \Pass$ be the closure under subset-formation of the set of $\vp(\alpha) \in \Pass$ for injections $\alpha: I \ra \tb 1k$ with $I \subset \tb 1m$ and $\abs{I} \le k-1$, where
\begin{displaymath}
  \vp(\alpha)(p_{i,j}) :=
  \begin{cases}
    1 & \text{if } i \in I \und j = \alpha(i)\\
    0 & \text{if } i \in I \und j \not= \alpha(j)\\
    0 & \text{if } \ex\, i' \in I : i' \not= i \und j = \alpha(i')\\
    \text{undefined} & \text{otherwise}
  \end{cases}.
\end{displaymath}
That is, $\vp(\alpha)$ sets $p_{i,j}$ to $1$ if $\alpha(i) = j$, and for these $i$ and $j$ makes sure that $i$ is not additionally mapped to some other $j'$, and that no other $i'$ is mapped to $j$. We show that $\PP$ is a $(k-1)$-consistent set of partial assignments for $\ofphp^m_k$, which by Theorem \ref{thm:methodwhdlb} shows the assertion of the theorem.

Consider $\vp \in \PP$ and a variable $p_{i,j} \in \var(\ofphp^m_k) \sm \var(\vp)$. Assume that for both $\ve \in \set{0,1}$ holds $\vp \cup \pao{p_{i,j}}{\ve} \notin \PP$. So there is $I \subset \tb 1m$, $\abs{I} = k-1$, and an injection $\alpha: I \ra \tb 1k$ with $\vp \sse \vp(\alpha)$, where $i \notin I$ and $j \notin \set{\alpha(i') : i' \in I}$. Now consider any $\psi \sse \vp$ with $n(\psi) < k-1$. By definition of $\PP$ we have $\psi \cup \pao{p_{i,j}}{0}, \psi \cup \pao{p_{i,j}}{1} \in \PP$. \Qed
\end{prf}

Now also strengthening Lemma 6.2 in \cite{Ku99b}:
\begin{corol}\label{cor:upofhpwhardness}
  For $m > k \ge 0$ and $F \in \set{\php^m_k,\fphp^m_k, \ophp^m_k}$ we have $\whardness(F) = \hardness(F) = k$.
\end{corol}
\begin{prf}
We have $\hardness(\php^m_k) = k$ by Lemma 6.2 in \cite{Ku99b}, and the assertion follows, since the w-hardness is at most the hardness, and adding clauses to unsatisfiable clause-sets can not increase the (w-)hardness. \Qed
\end{prf}

For $\fphp^m_k, \ophp^m_k$ we could have applied the result from \cite{GwynneKullmann2013GoodRepresentationsIII} about ``totally blocked'' clauses . By the same reasoning and the isomorphism between $\ofphp^m_k$ and $\ofphp^k_m$ we obtain:
\begin{corol}\label{cor:whdofphp}
  $\whardness(\ofphp^m_k) = \hardness(\ofphp^m_k) = \min(m,k)$ for $m \not= k$.
\end{corol}

By Lemma \ref{lem:spacetres} and Theorem \ref{thm:resspace} we obtain:
\begin{corol}\label{cor:spacephp}
  For $m > k \ge 0$ and $F \in \set{\php^m_k,\fphp^m_k,\ophp^m_k, \ofphp^m_k}$ we have $k \le \semspace(F) \le \treespace(F) = k+1$.
\end{corol}

\begin{quest}\label{que:spacephp}
  Can the slackness of $1$ for $\semspace(\php^m_k)$ in Corollary \ref{cor:spacephp} be removed? See Conjecture \ref{con:improvesslb}.
\end{quest}

Now also considering satisfiable cases:
\begin{thm}\label{thm:phpwhardness}
  For all $m, k \in \NNZ$ we have
  \begin{displaymath}
    \whardness(\php^m_k) = \hardness(\php^m_k) = \min(\max(m-1,0),k).
  \end{displaymath}
  XXX all other variants XXX
\end{thm}
\begin{prf}
  XXX
\end{prf}

\subsection{An extension of $\php$: EPHP}
\label{sec:ephp}

In \cite{Co76} it is shown that the pigeonhole clause-sets $\php^{n+1}_n \in \Usat$ have polynomial-size Extended Resolution (ER) refutations, that is, an extension $\php^{n+1}_n \subset \ephp_n \in \Cls$ has been described via Tseitin's extension rule, and $\ephp_n$ has a resolution refutation of polynomial size.

\begin{defi}\label{def:ephp}
  Consider $n \in \NN_0$. The \textbf{extended pigeon-hole formulas} are defined as
  \begin{displaymath}
    \bmm{\ephp_n} := \php^{n+1}_n \cup
    \bc_{\substack{l \in \tb3{n+1}\\
        i \in \tb 1{l-1}\\
        j \in \tb 1{l-2}}}
    \primec_0(q^{l-1}_{i,j} \lra (q^l_{i,j} \oder (q^l_{i,l-1} \und q^l_{l,j}))),
  \end{displaymath}
  where XXX the coefficients are not understandable XXX more care is needed XXX
  \begin{enumerate}
  \item for all $l \in \tb3{n}$, $i \in \tb 1{l-1}$ and $j \in \tb {1}{l-2}$ we have that $q^l_{i,j}$ is a new distinct variable;
  \item for all $i \in \tb1{n+1}$ and $j \in \tb1n$ we have that $q^{n+1}_{i,j} := p_{i,j}$.
  \end{enumerate}
\end{defi}
Remarks:
\begin{enumerate}
\item Explicitly, for all $l \in \tb3{n+1}$, $i \in \tb 1{l-1}$ and $j \in \tb 1{l-2}$ we have
  \begin{multline*}
    \primec_0(q^{l-1}_{i,j} \lra (q^l_{i,j} \oder (q^l_{i,l-1} \und q^l_{l,j}))) =\\
    \setb{ \set{q_{l-1,i,j},\ol{q_{l,i,j}}}, \set{q_{l-1,i,j},\ol{q_{l,i,l-1}},\ol{q_{l,l,j}}},\\
      \set{\ol{q_{l-1,i,j}},q_{l,i,j}, q_{l,i,l-1}}, \set{\ol{q_{l-1,i,j}},q_{l,i,j},q_{l,l,j}} }.
  \end{multline*}
\item Definition \ref{def:ephp} provides the full extension XXX what does ``full'' mean here? XXX, for all levels $l \in \tb3{n+1}$ XXX what is special about ``all levels'' ? XXX, for the extended resolution proof in \cite{Co76} of $\php^{n+1}_n$.
\item we need full details here XXX at least precise details about the sizes XXX the resolution refutation provided in the appendix XXX
\item Definition \ref{def:ephp} is equivalent to $\ephp^{n+1}_n$ in \cite{JarvisaloJunttila2009LimitRestrictedLearning}, modulo the removal of additional variables occurring in \cite{JarvisaloJunttila2009LimitRestrictedLearning} due to using only a restricted extension (recall Definition \ref{def:exresproof}). The redundant variables are immediately removed in the short resolution proof of $\ephp_n$ given in the appendix of \cite{JarvisaloJunttila2009LimitRestrictedLearning}, and essentially the same resolution proof applies to our form of $\ephp_n$. XXX also comparison to the EPHP refutation in \cite{Ku96c}
\end{enumerate}

\begin{thm}\label{thm:hdephp}
  For $n \in \NN$ holds $\hardness(\ephp_n) = n$ XXX
\end{thm}
\begin{prf}
  XXX
\end{prf}

 So $\ephp_n$ is not useful for tree-resolution based SAT-solvers (the core of ``look-ahead solvers''; see \cite{HvM09HBSAT}), and can possibly only be exploited by conflict-driven SAT solvers (see \cite{MSLM09HBSAT}); this answers the open question on the tree-resolution complexity of the extended pigeon-hole formulas posed in \cite{Jarvisalo2011EfficiencyDPLLOBDD}.

\begin{conj}\label{con:whdephp}
  For $n \in \NN$ holds $\whardness(\ephp_n) = \hardness(\ephp_n)$.
\end{conj}

\begin{quest}\label{que:ephprmbl2}
  Let $\ephp_n'$ be the result of removing iteratively all blocked 2-clauses from $\ephp_n$ ($n \in \NNZ$).
  \begin{enumerate}
  \item We have $c(\ephp_n) - c(\ephp_n') = n \cdot (n+1)/2-1$.
  \item Do we have an exponential resolution lower bound for $\ephp_n'$?
  \item The question is whether the result from \cite{Ku96c}, that ``blocked $2$-extensions'' can not be simulated polynomially by resolution, can be sharpened so that addition of blocked clauses can not simulated? (The point is that \cite{Ku96c} allows addition of clauses iff they become blocked in \emph{some} removal order.)
  \end{enumerate}

\end{quest}

\subsection{Tree resolution and ER}
\label{sec:treeresER}

Basically rephrasing \cite{JarvisaloJunttila2009LimitRestrictedLearning,Jarvisalo2011EfficiencyDPLLOBDD} (XXX details needed):
\begin{lem}\label{lem:full2treevisext}
  Consider $F \in \Usat$ and a resolution refutation $R$ for $F$ (see Definition \ref{def:resproof}). Let
  \begin{displaymath}
    E(R) := \bc_{C \in \allcr(R) \sm (F \cup \set{\bot})} \primec_0(e_C \lra C).
  \end{displaymath}
  where the $e_C$ are new variable (one for each different clause in $R$). Then we have:
  \begin{enumerate}
  \item $F' := F \cup E(R)$ is an extension of $F$ (see Definition \ref{def:exresproof}).
  \item $\hardness(F') \le 2$.
  \end{enumerate}
\end{lem}
\begin{prf}
  That we have an extension follows by definition. To show $\hardness(F') \le 2$, we show that $\rk_2$ for input $F'$ sets all variables $e_C$ to true, from which the assertion follows, since then we obtain $\set{v}, \set{\ol{v}}$ for some variable $v$. Set $e_C \ra 0$. Then all literals of $C$ become $0$ via $\rk_1$. Now the two parent clauses of $C$ are present in $F'$, since either one of the parent clauses $D$ is an axiom, or inductively already $e_D \ra 1$ was set. So we obtain a contradiction via $\rk_1$. \Qed
\end{prf}

A special case of Lemma \ref{lem:full2treevisext} is the construction in \cite{JarvisaloJunttila2009LimitRestrictedLearning} of $\ephp_n \subset \ephp_n'$, for which we got now $\hardness(\ephp_n') \le 2$, and thus this extension of $\php^{n+1}_n$ now is really easy also for tree-based SAT solvers. XXX

\subsection{Monotone circuits for $\php^m_m$}
\label{sec:phpmonotone}

By Theorem \ref{thm:phpwhardness} we know that the w-hardness of $\php^m_m$ is unbounded. Could there be clause-sets $B_m$ equivalent to $\php_m^m$ with bounded hardness? Such questions are relevant for SAT solving, since with a ``reasonable'' $B_m$ we could express the bijectivity condition of $\php^m_m$, when needed as part of a SAT problem, in a form better ``understandable'' than by $\php^m_m$, that is, producing not such hard sub-instances via instantiation (and instantiation is precisely the business of a SAT solver). Of course there is an equivalent clause-set in $\Urefc_0$, namely $\primec_0(\php^m_m)$ (this is also essentially unique, up to subsumption), which however is of exponential size (recall Lemma \ref{lem:prcpig}). So we require that the size of $B_m$ is polynomial in $m$. We will see in this subsection that there are no such $B_m$, even if we allow $B_m$ to contain auxiliary variables like in EPHP, and even if we restrict the defining condition of w-hardness for $B_m$ to the variables in $\php^m_m$, that is ignoring the auxiliary variables (using relative hardness, as introduced in Subsection \ref{sec:extmeasures}).

In \cite{GwynneKullmann2013GoodRepresentationsII} the following theorem has been shown (motivated by a similar result in \cite{BKNW2009CircuitComplexity}; recall that a monotone circuit only uses binary and's and or's):
\begin{thm}\label{thm:mononII}
  For a boolean function $f(v_1,\dots,n)$ (in $n$ variables) the monotonisation $\widehat{f}(v_1',v_1,'',\dots,v_n',v_n'')$ (in $2n$ variables) is defined by $\widehat{f}(v_1',v_1,'',\dots,v_n',v_n'')=1$ iff there is a vector $(v_1,\dots,v_n) \in \set{0,1}^n$ with $f(v_1,\dots,v_n) = 1$, such that for all $i \in \tb 1n$ holds: if $v_i = 0$, then $v_i'=1$, and if $v_i = 1$, then $v_i''=1$.

  Consider a clause-set $F \in \Cls$ with $\set{v_1,\dots,v_n} \sse \var(F)$, such that $F$ ``represents'' $f$ in the sense that the satisfying assignments of $F$ projected to $\set{v_1,\dots,v_n}$ are precisely the satisfying assignments of $f$. Now from $F$ in cubic time a monotone circuit $\mc{C}$ in inputs $v_1',v_1,'',\dots,v_n',v_n''$ can be constructed, such that $\hardness^{\set{v_1,\dots,v_n}}(F) \le 1$ if and only if $\mc{C}$ computes $\widehat{f}(v_1',v_1,'',\dots,v_n',v_n'')$.
\end{thm}
Let $f_m$ be the boolean function of $\php^m_m$, with variables $p_{i,j}$, $i, j \in \tb 1m$, and consider the monotonisation $\widehat{f_m}$, with variables $p_{i,j}', p_{i,j}''$, $i, j \in \tb 1m$. Furthermore let the boolean function $g_m$ on variables $p_{i,j}$, $i, j \in \tb 1m$, be the perfect-matching function, that is, $g_m$ is true iff the bipartite graph given by the edges $\set{i,j}$ with $p_{i,j}=1$ contains a perfect matching. Via $p_{i,j}':=1, p_{i,j}'':=p_{i,j}$ we can compute $g_m$ from $\widehat{f_m}$. Razborov's lower bound (see Theorem 9.38 in \cite{Jukna2012BooleanFunctionComplexity} for a nice presentation) says that every monotone circuit computing $g_m$ has $m^{\Omega(\log m)}$ gates, and thus also $\widehat{f_m}$ has $m^{\Omega(\log m)}$ gates. Thus we have shown:
\begin{thm}\label{thm:repphp}
  If $F$ is a representation of $\php^m_m$ with $\hardness^{\set{p_{i,j}}}(F) \le 1$, then $\ell(F) = m^{\Omega(\log m)}$.
\end{thm}
A relevant question is how sharp the bound of Theorem \ref{thm:repphp} is; the current bound leaves a practical potential for such $F$, since for practical applications it seems $m \le 100$ can be assumed. In \cite{GwynneKullmann2013GoodRepresentationsIII} it is shown that $\whardness^V \le k$ for fixed $k$ can be transformed in polynomial time to $\hardness^V \le 1$ (when allowing auxiliary variables), and thus we get:
\begin{thm}\label{thm:repphpgen}
  For every $k \in \NNZ$ the size of representations $F$ of $\php^m_m$ with $\whardness^{\set{p_{i,j}}}(F) \le k$ grows superpolynomially in $m$.
\end{thm}
Again the question is how small such representations in dependency on $k$ could be (for practically relevant values of $k$ there could be interesting representations). Also of relevance to ask about the sizes of representations with given absolute (w-)hardness, i.e., representations $F$ of $\php^m_m$ with $\hardness(F) \le k$ resp.\ $\whardness(F) \le k$.

\section{Application: XOR}
\label{sec:expxor}

\subsection{Simple example: Two equations}
\label{sec:xor2eq}

We consider the representations $X_0, X_1: \Cls \ra \Cls$ of XOR-clause-sets $F$ via CNF-clause-sets $X_0(F), X_1(F)$ as investigated in \cite{GwynneKullmann2013GoodRepresentationsII}.

\begin{conj}\label{con:unbhd2xc}
  There are $F \in \Cls$ with $c(F) = 2$ and $\hardness(X_1(F^*))$ arbitrarily large (using $F^*$ as defined in \cite{GwynneKullmann2013GoodRepresentationsII}, namely $F^* = \set{\oplus F' : F' \sse F}$).
\end{conj}

\begin{thm}\label{thm:2xor}
  For $n \in \NN$ and (different) variables $v_1,\dots,v_n$ consider the system
  \begin{eqnarray*}
    v_1 \oplus v_2 \oplus \dots \oplus v_n &=& 0 \\
    v_1 \oplus v_2 \oplus \dots \oplus \ol{v_n} &=& 0,
  \end{eqnarray*}
  that is, consider the XOR-clauses $C_1 := \set{v_1,\dots,v_n}$ and $C_2 := \set{v_1,\dots,v_{n-1},\ol{v_n}}$. First we remark that $X_0(\set{C_1,C_2})$ is the clause-set with all $2^n$ full clauses of $\set{v_1,\dots,v_n}$, and thus $\hardness(X_0(\set{C_1,C_2})) = \whardness(X_0(\set{C_1,C_2})) = n$. Now let $T_n := X_1(\set{C_1,C_2})$ (see XXX). We have $\hardness(F) = n$. XXX While in XXX it is shown that $\wid(F) = 3$, and indeed $F$ has linear size resolution refutations.
\end{thm}
\begin{prf}
From Corollary \ref{cor:hdunion} and Lemma XXX (in PC) we obtain $\hardness(T_n) \le n+1$. Better is to apply Lemma \ref{lem:hdpassub} with $V := \set{v_2,\dots,v_{n-1}}$. By definition we see that $\psi * T_n \in \Pcls{2}$ (i.e., all clauses have length at most two) for $\psi$ with $\var(\psi) = V$. By Lemma 19 in \cite{GwynneKullmann2012SlurSOFSEM} we have $\hardness(\psi * T_n) \le 2$, and thus $\hardness(T_n) \le (n-2) + 2 = n$.

The lower bound is obtained by an application of Lemma \ref{lem:lbhd}. Consider any literal $x \in \lit(T_n)$. Setting $x$ to true or false results either in an equivalence or in an anti-equivalence. Propagating this (anti-)equivalence yields a clause-set $T'$ isomorphic to $T_{n-1}$, where by Lemma \ref{lem:eqhardness} this propagation does not increase hardness, so we have $\hardness(\pao{x}{1} * T_n) \ge \hardness(T') = \hardness(T_{n-1})$. The argumentation can be trivially extended for $n \in \set{0,1,2}$, and so by Lemma \ref{lem:lbhd} we get $\hardness(T_n) \ge n$. XXX \Qed
\end{prf}

\begin{corol}\label{cor:x1two}
  For two XOR-clauses $C,D \in \Cl$ except of trivial exceptions XXX holds
  \begin{displaymath}
    \hardness(X_1(\set{C,D})) =
    \hardness^{\var(\set{C,D})}(X_1(\set{C,D})) = \max(1,\abs{\var(C)
      \cap \var(D)}).
  \end{displaymath}
\end{corol}
\begin{prf}
  XXX
\end{prf}

\begin{corol}\label{cor:hdTs}
  The Tseitin translation, applied to a boolean circuit, has unbounded hardness in general, for the full form as well as the reduced form, as can be seen by the circuit computing via binary xor's in two chains the two sums $v_1 \oplus \dots \oplus v_n$ and $v_1 \oplus \dots \oplus \ol{v_n}$, and where the final circuit, computing the (single) output, is the equivalence of these two sums: The full Tseitin translation has hardness $n$ by Theorem \ref{thm:2xor}, and thus also the reduced Tseitin translation, which yields an (unsatisfiable) sub-clause-set, has hardness at least $n$.
\end{corol}

\begin{lem}\label{lem:unbhd2xc}
  For two XOR-clauses $C, D$, $\hardness(X_1(\set{C,D,\oplus\set{C,D}}))$ is arbitrarily large.
\end{lem}
\begin{prf}
  XXX
\end{prf}

\subsection{Tseitin clause-sets}
\label{sec:Tseitincls}

A ``hypergraph'' is a pair $G = (V,E)$, where $V$ is a set and $E$ is a set of finite subsets of $V$; one writes $V(G) = V$ and $E(G) = E$. A ``general hypergraph'' is a triple $(V,E,e)$, where $V, E$ are sets and $e: E \ra \pote(V)$, where $\pote(X)$ for a set $X$ is the set of finite subsets of $X$; one writes $e(G) = e$.

An ``XOR-constraint'', or a linear equation over $\ZZ_2$, is a finite set $V \subset \Va$ of variables together with $\ve \in \set{0,1}$, with the interpretation ``$\oplus_{v \in V} v = \ve$''. So a \emph{system of XOR-constraints/linear equations} is a pair $(G,\rho)$, where $G$ is a finite hypergraph with $V(G) \sse \Va$, and $\rho: E(G) \ra \set{0,1}$ assigns to each hyperedge (an equation) the prescribed sum. The basic associated clause-set is $\bmm{X_0(G,\rho)} \in \Cls$ defined as
\begin{displaymath}
  X_0(G,\rho) := X_0(\set{\oplus_{v \in H} = \rho(H)}_{H \in E(G)}).
\end{displaymath}

The \textbf{dual} of $(G,\rho)$, written $\trans{(G,\rho)}$, is the pair $(\trans{G}, \rho)$, where
\begin{itemize}
\item $\trans{G}$ is the dual of $G$ as general hypergraph, that is:
  \begin{itemize}
  \item $V(\trans{G}) = E(G)$
  \item $E(\trans{G}) = V(G)$
  \item the hyperedge-function $e: E(\trans{G}) \ra \pote(V(\trans{G}))$ assigns to every $v \in V(G)$ the set of $H \in E(G)$ with $v \in H$;
  \end{itemize}
\item so now $\rho: V(\trans{G}) \ra \set{0,1}$.
\end{itemize}
In general, a \emph{dual system of XOR-constraints/linear equations over $\ZZ_2$} is a pair $(G,\rho)$, where $G$ is a finite general hypergraph with $E(G) \sse \Va$ and $\rho: V(G) \ra \set{0,1}$. So the associated system of XOR-constraints is obtained again by dualisation, written again $\trans{(G,\rho)} := (\trans{G},\rho)$, where $\trans{G}$ is the dual of $G$ as (ordinary) hypergraph, that is, $V(\trans{G}) = E(G)$ and $E(\trans{G}) = \set{v \in E(G) : w \in e(G)(v)}_{w \in V(G)}$. The associated clause-set $\bmm{X_0(G,\rho)} \in \Cls$ is $X_0(G,\rho) := X_0(\trans{(G,\rho)})$.

Obviously dualisation in both directions yields inverse bijections between the set of systems of XOR-constraints and the set of dual systems of XOR-constraints.

A \textbf{full Tseitin graph} is a dual system of XOR-constraints $(G,\rho)$, where $G$ is a connected irreflexive general graph with $\oplus_{w \in V(G)} \rho(w) = 1$, where irreflexive general graphs says $\fa\, v \in E(G) : \abs{e(G)(v)} = 2$. Note that additionally to ordinary (full) Tseitin graphs we allow parallel edges, but still loops are disallowed (a loop at a vertex in effect deactivates the corresponding equation). Now $X_0(G,\rho) \in \Usat$.

An important abstraction is obtained by the insight, that $X_0(G,\rho)$ and $X_0(G,\rho')$ are flipping-isomorphic, that is, by flipping literals we can obtain the former from the latter. So we consider plain connected irreflexive general graph with at least one vertex as \textbf{Tseitin graphs}, considering implicitly the set of all possible vertex-labellings (with elements from $\set{0,1}$, so that the (XOR-)sum is $1$).

To understand $\hardness(X_0(G))$ and $\whardness(X_0(G))$, we need to understand what splitting does with $G$. The variables $v \in \var(F)$ of $F := X_0(G)$ are the edges of $G$:
\begin{itemize}
\item If $G' := G - v$ is still connected, then $\pao v0 * F$ and $\pao v1 * F$ are both isomorphic to $X_0(G')$. Note that $G'$ is still a Tseitin graph.
\item Otherwise let $G', G''$ be the connected components of $G$ (both again Tseitin graphs). Now $\pao v0 * F$ and $\pao v1 * F$ are isomorphic, in some order, to $X_0(G'), X_0(G'')$.
\end{itemize}
The endpoint of splitting is reached when $G$ is the one-point graph (which can not have edges, since loops are not allowed). So we can formulate the hardness and w-hardness games for Tseitin graphs:

\begin{lem}\label{lem:charachdTs}
  Let $G$ be a Tseitin graph. Then $\hardness(X_0(G))$ is characterised by the following game:
  \begin{enumerate}
  \item An atomic move for the current non-trivial Tseitin graph $G$ replaces $G$ with a sub-graph $G'$ of $G$, obtained by choosing some $e \in E(G)$ and choosing a connected component of $G - e$.
  \item The two players play in turns, and delayer starts with $G$.
  \item A move of delayer is to apply a sequence of atomic moves (possibly zero).
  \item A move of prover is to apply one atomic move.
  \item The games ends when $G$ becomes trivial, in which case delayer gets as many points as there have been moves by prover.
  \end{enumerate}
\end{lem}

\begin{lem}\label{lem:characwhdTs}
  Let $G$ be a Tseitin graph. Then $\whardness(X_0(G))$ is characterised by the following game:
  \begin{enumerate}
  \item The notion of ``atomic move'' is as in Lemma \ref{lem:charachdTs}.
  \item Again, two players play in turns, and delayer starts with $G$.
  \item Again, a move of delayer is to apply a sequence of atomic moves (possibly zero).
  \item A mover of prover is to replace the current global sequence of atomic moves by another sequence, which is consistent with the old sequence, and handles exactly one new edge.
  \item Here ``consistent'' means that in case removal of an edge splits the graph into two connected components, where this edge occurs also in the original sequence, then the same ``side'' of the graph is chosen.
  \item The games ends when $G$ becomes trivial, in which case delayer gets as many points as the maximum length of a sequence used in a replacement by prover.
  \end{enumerate}
\end{lem}

\section{Conclusion and outlook}
\label{sec:conclusion}

\paragraph{Acknowledgements}

I thank \href{http://cs.swan.ac.uk/~csmg/}{Matthew Gwynne} for fruitful discussions.

\bibliographystyle{plain}

\newcommand{\noopsort}[1]{}

\end{document}